\newif\ifmsinclaude
\makeatletter
\IfFileExists{../main.tex}{\msinclaudetrue\def\input@path{{./}{../}}}{\def\input@path{{./}{Claude_v2/}}}
\makeatother
\newif\ifmsbibinclaude
\ifmsinclaude
  \edef\msjobname{\jobname}\edef\msoverleafjob{\detokenize{output}}
  \ifx\msjobname\msoverleafjob\else\msbibinclaudetrue\fi
\fi
\documentclass[a4paper,11pt]{article}
\usepackage[utf8]{inputenc}
\usepackage[T1]{fontenc}
\usepackage{jcappub}
\usepackage{xcolor}
\graphicspath{{./}{../}}
\usepackage{xurl}

\title{Cosmic Ray Heating of the Early IGM: Realistic Galaxy Populations, Reduced Efficiency, and 21-cm Signatures}

\author[a,b,1]{Michael Pagano,\note{Corresponding author.}}
\author[c]{Andrei Mesinger,}
\author[a,b]{Carmelo Evoli}
\author[d]{and Sruthi Ravikularaman}

\affiliation[a]{Gran Sasso Science Institute (GSSI), Viale Francesco Crispi 7, 67100 L'Aquila, Italy}
\affiliation[b]{INFN-Laboratori Nazionali del Gran Sasso (LNGS), via G. Acitelli 22, 67100 Assergi (AQ), Italy}
\affiliation[c]{Department of Physics and Astronomy {\it ``Ettore Majorana''}, University of Catania, Via Santa Sofia 64, 95123 Catania, Italy}
\affiliation[d]{Universit\'e Paris Cit\'e, CNRS, Astroparticule et Cosmologie, F-75013 Paris, France}

\emailAdd{michael.pagano@gssi.it}

\abstract{Cosmic rays (CRs) accelerated by the first supernovae could heat the intergalactic medium (IGM) before reionization. Previous 21-cm simulations, assuming a kinetic-energy power-law spectrum and simplified galaxy populations, found that CR heating can rival X-ray heating from high-mass X-ray binaries, imprint a distinct signature on the 21-cm power spectrum, and flatten the absorption trough of the global 21-cm signal. Here we model CR heating in \texttt{21cmFAST} with: (i) the momentum power-law spectrum predicted by diffusive shock acceleration; (ii) Coulomb losses off free electrons; and (iii) injection from data-informed populations of atomic-cooling and molecularly cooled galaxies. We find that CR heating is about an order of magnitude weaker than previously estimated because a momentum power law places little energy in the sub-relativistic protons ($\lesssim40$\,MeV) that supply 99\% of the heat. For our fiducial spectral slope, at the hard end of those inferred for Galactic supernova remnants, and for matched efficiency, this reduces the heating by a factor of $\approx12$, and together with a more realistic galaxy model, by a factor of $\approx28$ at $z = 15$. At the efficiencies that supernovae can supply, CR heating is weaker than standard X-ray heating, and the global signal alone cannot tell the two apart. Distinguishing CR from X-ray heating with the 21-cm power spectrum requires CR efficiencies a few times higher than those expected from local observations. At such efficiencies, CR heating yields up to $\approx55\%$ more power than X-ray heating at matched global signal (median over scales), compared with only 3--6\% at a locally motivated efficiency. A flattened absorption trough, like the one claimed by EDGES, requires efficiencies over two orders of magnitude above the local ones, beyond plausible Population~III supernova budgets.}


\begin{document}
\maketitle
\flushbottom
\label{firstpage}

% ============================================================
\section{Introduction}\label{sec:intro}

The redshifted 21-cm line of neutral hydrogen opens an observational window on the intergalactic medium (IGM) through the dark ages, cosmic dawn and the epoch of reionization~\citep{Furlanetto2006review, PritchardLoeb2012}. Both the sky-averaged global signal and its spatial power spectrum encode the processes that shape the IGM, including structure formation, ionization, Lyman-$\alpha$ (Ly$\alpha$) coupling, and the heating of the gas after it has thermally decoupled from the cosmic microwave background (CMB). A concerted experimental effort, including HERA~\citep{DeBoer2017}, LOFAR~\citep{vanHaarlem2013}, the MWA~\citep{Tingay13}, NenuFAR~\citep{Zarka2012} and, in the future, the SKA~\citep{Koopmans2015}, is now producing upper limits on the 21-cm power spectrum (e.g.~\citep{LOFAR25, Nunhokee2025, NANUFAR25}) that are beginning to constrain astrophysical models. The tightest, from HERA, already exclude the most extreme cold-reionization scenarios~\citep{HERA2022a, HERA2022b, HERA2023}, and its first Phase~II results extend the limits to cosmic-dawn redshifts~\citep{HERA2025}.

The excluded scenarios are those in which the IGM stays cold. Once Wouthuysen--Field coupling drives the spin temperature toward the kinetic temperature of the gas, the contrast of the 21-cm signal against the CMB tracks the IGM temperature (section~\ref{sec:21cm}), making the signal a thermometer of the first heat sources.\footnote{Non-standard physics can alter this picture in either direction. Scattering between baryons and dark matter can cool the gas below the adiabatic track and deepen the absorption~\citep{Barkana2018, MunozLoeb2018}, while energy injection from annihilating or decaying dark matter can heat the IGM earlier and more uniformly than astrophysical X-ray sources~\citep{Evoli2014, LiuSlatyer2018}.} X-rays from high-mass X-ray binaries and the hot interstellar medium (ISM) of early galaxies have long been regarded as the dominant heat source~\citep{Furlanetto2006heating, PF07, Mesinger2013, Pacucci2014, Fialkov2014}. Their population-averaged luminosity per unit star formation rate (SFR), $L_{\rm X}/{\rm SFR}$, is poorly constrained at high redshift: locally calibrated relations already span more than an order of magnitude and depend systematically on metallicity~\citep{MGS12_HMXB, Brorby2016, Lehmer2021}, and the empirical calibrations reach only $12+\log({\rm O/H}) \approx 7$~\citep{Lehmer2021}, below which the relation rests on binary population synthesis~\citep{Fragos2013}. This freedom drives the degeneracies that currently limit the astrophysical conclusions drawn from 21-cm upper limits.

The massive stars whose remnants power these X-ray binaries also end their lives as core-collapse supernovae, whose explosions accelerate cosmic rays (CRs). Diffusive shock acceleration (DSA) at supernova-remnant shocks~\citep{Bell1978, BlandfordOstriker1978} produces relativistic protons and, subdominantly, leptons, channeling an estimated 5--30\% of the explosion energy into CRs~\citep{Vink2012, Caprioli2012, Morlino2012}. Locally, CRs play an important role in regulating star formation and in the thermodynamics of the ISM and circumgalactic medium~\citep{Padovani2020, Ruszkowski2023}. If they escape their host galaxies in sufficient numbers, they can heat the neutral and partially ionized IGM through the excitation and ionization of atomic hydrogen and through Coulomb scattering off free electrons~\citep{SazonovSunyaev2015, Leite2017, JanaNath2018}. How many escape depends on the unknown magnetic-field geometry of the first galaxies, a dependence that is usually absorbed into a single efficiency parameter.

The thermal imprint of CR heating on the 21-cm signal has been studied by several authors. Following early work on CR heating of the IGM by young galaxies~\citep{Nath1993}, \citet{SazonovSunyaev2015} argued that CRs from primordial supernovae could preheat the IGM ahead of reionization, and proposed the escape channel from molecularly cooled galaxies (MCGs) that we take up in section~\ref{sec:direct_injection_physics}. \citet{Leite2017} computed the resulting thermal history in detail, for a momentum power-law injection spectrum and including Coulomb losses off free electrons, and found that CRs can heat the IGM by $\Delta T \approx 10$--$200$\,K by $z = 10$, depending strongly on the CR injection spectrum. \citet{JanaNath2018} followed the diffusion of low-energy CRs around individual high-redshift halos and found the heating to be nearly uniform for typical diffusion coefficients. \citet{Bera2023} computed the global 21-cm signal for CRs injected by both Population~III (Pop~III) and Population~II (Pop~II) supernovae, finding that the depth, position and duration of the absorption trough all respond to the assumed CR efficiencies, and \citet{YokoyamaOhira2023} identified a further channel, in which the return current induced by streaming CRs heats the gas resistively, dominating close to the source galaxy. None of these studies computed the spatially resolved 21-cm signal: \citet{Leite2017} and \citet{YokoyamaOhira2023} noted that heating clustered around the sources would leave a signature in 21-cm maps, but computed the thermal history only for the volume-averaged IGM or around a single source.

\citet{GesseyJones2023} (hereafter GJ23) investigated the spatially resolved 21-cm signatures of CR heating with semi-numerical simulations, making three claims that motivate this work. First, CR heating can compete with X-ray heating: at the standard efficiencies of the literature, they found CRs roughly as efficient as X-rays at heating the neutral IGM at $z<20$, and slightly more efficient at raising its temperature. Second, CR heating is short-ranged: the protons deposit their energy close to their sources, so the temperature field retains small-scale structure that X-ray heating smooths away, amplifying the 21-cm power spectrum at high wavenumbers and offering a way to identify the dominant heating mechanism with a future detection. Third, if CRs escape efficiently only from MCGs, the rising Lyman--Werner (LW) background switches off CR injection by suppressing star formation in the smallest halos. This produces a local peak in the IGM temperature and, for a limited parameter range, a flattened absorption trough in the global signal, a morphology reminiscent of the profile claimed by EDGES~\citep{Bowman2018} (though see~\citep{Singh2022}).

The GJ23 models, however, rest on simplified ingredients. They parametrize the CR injection spectrum as a power law in kinetic energy, whereas DSA predicts a power law in momentum; they assume a constant ratio of stellar to halo mass, whereas the UV luminosity functions of high-redshift galaxies favor a ratio that decreases toward low halo masses (section~\ref{sec:galaxies}); and they neglect Coulomb losses in the partially ionized IGM outside ionized regions. The injection spectrum and the galaxy model together set how much energy is in the sub-relativistic protons that heat the IGM, and when this heating occurs. It is therefore not obvious that the three claims survive a physically motivated injection spectrum and more observationally motivated galaxy populations.

Here we revisit these claims with a new implementation of CR heating in the semi-numerical code \texttt{21cmFAST}~\citep{MF07, Mesinger2011, Murray2020}. Specifically, we: (i) replace the kinetic-energy power-law injection spectrum with the momentum power law predicted by DSA, which differs from it most at the sub-relativistic energies that dominate IGM heating; (ii) include Coulomb losses off free electrons~\citep{Schlickeiser2002, Leite2017} alongside the excitation, ionization and adiabatic losses; and (iii) source the CRs from the data-informed galaxy populations of the EOS21 model~\citep{Munoz2022}, with a mass-dependent stellar-to-halo mass relation for Pop~II stars in atomic-cooling galaxies (ACGs) and separate injection from Pop~III stars in MCGs. We then compare CR and X-ray heating in the thermal history, the global 21-cm signal and the 21-cm power spectrum, and revisit MCG-only injection and the plausibility of a flattened global absorption trough.

%Several findings emerge from our work. Our momentum power-law spectrum, evaluated at parameters matched to \citet{GesseyJones2023}, produces less heating at $z \sim 15$. This reduction tightens the degeneracy between CR and X-ray heating at the level of the thermal history; for a given $\eta_{\rm cr}$, the resulting $T_K(z)$ is closer to what moderately optimistic X-ray heating produces. Coulomb losses off free electrons, included here but neglected in earlier 21-cm CR studies, contribute at the few-percent level to the thermal history and leave the 21-cm observables essentially unchanged. Including CR heating as an additional channel can broaden the posterior on $L_X/\mathrm{SFR}$ inferred from HERA upper limits, tempering the most aggressive conclusions one might otherwise draw from an X-ray-only analysis of the first high-mass X-ray binaries. We leave this to future work. Because the thermal history alone cannot discriminate between the two channels, distinguishability in the detection era will rely on the $k$-dependent shape of the 21-cm power spectrum, where the different mean free paths of CRs and X-ray photons leave distinct spatial signatures \citep{Leite2017, JanaNath2018, GesseyJones2023}.

This paper is organized as follows. In section~\ref{sec:physics} we review the energy losses of CR protons, derive the momentum power-law injection spectrum, and estimate the range of injection efficiencies that supernovae can supply. In section~\ref{sec:implementation} we describe the galaxy populations, the implementation of CR transport and heating in \texttt{21cmFAST}, MCG-only injection, the 21-cm observables and our simulations. In section~\ref{sec:Results} we identify which protons heat the IGM, compare CR and X-ray heating in the thermal history, quantify how the injection spectrum and the galaxy model set the heating relative to GJ23, compare the two channels in the global signal and the power spectrum, and examine MCG-only injection. We conclude in section~\ref{sec:conclusions}. Our cosmological parameters are consistent with the Planck 2018 results~\citep{Planck2018}, and our astrophysical parameters are those of the EOS21 model of \citet{Munoz2022} unless stated otherwise.

% ============================================================
\section{Cosmic-ray heating physics}
\label{sec:physics}

The CR heating rate of the IGM is set by three ingredients, which we describe in turn: the energy losses of CR protons and the fraction of each that heats the gas (section~\ref{sec:losses}), the injection spectrum (section~\ref{sec:spectrum}), and the CR energy injected per unit stellar mass formed (section~\ref{sec:escape}).

\subsection{Energy-loss channels}
\label{sec:losses}

Supernova remnants accelerate both electrons and ions, but the CR energy budget is expected to be carried mainly by protons. CR electrons carry a small fraction of the injected energy and, at high redshift, lose it through inverse Compton scattering off the CMB, synchrotron emission and bremsstrahlung rather than through collisional heating. Following earlier studies \citep{SazonovSunyaev2015,Leite2017,JanaNath2018}, we therefore consider only CR protons.

A CR proton propagating through the (partially) neutral IGM loses energy through three channels, which differ in how much of the lost energy heats the gas. (i) Hubble expansion redshifts the proton's momentum; these adiabatic losses dominate at relativistic energies but do not heat the gas. (ii) Excitation and ionization of neutral hydrogen and helium \citep{SazonovSunyaev2015, JanaNath2018} dominate the losses below $E_{\rm K} \approx 5$--$10\,\mathrm{MeV}$ at $z = 10$--$20$, the energies that supply almost all of the IGM heating (section~\ref{sec:attribution}). As the ejected electrons share their energy through a cascade of secondary ionizations and excitations, only a fraction $f_{\rm heat}(x_{\rm e})$ of it is deposited as heat. (iii) Coulomb scattering off free electrons \citep{Schlickeiser2002, Leite2017} grows in importance as the gas is ionized, and heats the gas in full. Hadronic, synchrotron and inverse Compton losses are negligible for the sub-GeV protons that dominate the heating \citep{Schlickeiser2002}.

The adiabatic loss due to cosmic expansion is
\begin{equation}
\left.\frac{\mathrm{d}E_{\rm K}}{\mathrm{d}z}\right|_{\rm ad}
= \frac{E_{\rm K}}{1+z}\left(2 - \frac{1}{1 + m_{\rm p} c^{2}/E_{\rm K}}\right),
\label{eq:hubble_loss}
\end{equation}
where $m_{\rm p}$ is the proton mass. Being proportional to $E_{\rm K}$, it dominates the losses of relativistic protons, for which the collisional channels are comparatively weak.

Excitation and ionization proceed through the Coulomb interaction between the CR proton and the bound electrons of the atoms it passes. Following \citet[eq.~(4.32)]{Mannheim1994} (see also~\citep{Schlickeiser2002}), we write the resulting loss rate, given by the Bethe--Bloch formula, as
\begin{equation}
\begin{split}
\left.\frac{\mathrm{d}E_{\rm K}}{\mathrm{d}t}\right|_{\rm E\&I}
&= -1.82 \times 10^{-7}\,\mathrm{eV\,s^{-1}\,cm^3}\,
\left(n_{\rm HI} + 2\,n_{\rm HeI}\right) \\
&\times \left[1 + 0.0185 \ln\beta\,\Theta(\beta - \beta_0)\right]
\frac{2\beta^2}{\beta_0^3 + 2\beta^3},
\end{split}
\label{eq:bethe_bloch}
\end{equation}
where $\beta = v/c$ is the proton velocity, $n_{\rm HI}$ and $n_{\rm HeI}$ are the number densities of neutral hydrogen and helium (the factor of two counting the two bound electrons of helium), $\Theta$ is the Heaviside step function, and $\beta_0 = 0.01$ is the characteristic orbital velocity of the bound electrons, of order the fine-structure constant. Below $\beta_0$, the proton is slower than the electrons it would ionize and the loss rate turns over; above it, the logarithmic term is a correction of at most $9\%$ that vanishes as $\beta \to 1$, where the loss rate tends to a constant. We assume that helium is as neutral as hydrogen, and omit the molecular-hydrogen term, $2 n_{\rm H_2}$, as H$_2$ is negligible in the diffuse IGM. The loss rate thus peaks (the Bragg peak) at $\beta = \beta_0$, i.e., at $E_{\rm K} \approx 47\,\mathrm{keV}$, and falls steeply toward higher energies.

This energy dependence has two consequences that are central to this work. First, it is sub-relativistic protons that deposit the bulk of their energy as heat, and their stopping distances are short: propagating ballistically, a $1\,\mathrm{MeV}$ ($0.1\,\mathrm{MeV}$) proton at $z = 10$ thermalizes within $\approx 2$ ($\approx 0.04$)\,cMpc, far less than the mean free path of the X-ray photons that carry the competing heating channel,
\begin{equation}
\lambda_{\rm X} \approx 20\,\langle x_{\rm HI}\rangle^{-1}
\left(\frac{E_{\rm X}}{300\,\mathrm{eV}}\right)^{2.6}
\left(\frac{1+z}{10}\right)^{-2}\,\mathrm{cMpc},
\label{eq:lambda_x}
\end{equation}
where $\langle x_{\rm HI}\rangle$ is the mean neutral fraction and $E_{\rm X}$ the photon energy \citep{Mesinger2013, Mesinger2015SKA}; at $z = 10$, a $1\,\mathrm{keV}$ photon travels $\sim 400\,\mathrm{cMpc}$ before being absorbed. CR heating is therefore expected to be more tightly correlated with the sources than X-ray heating (though see~\citep{JanaNath2018} for the effect of diffusion, and section~\ref{sec:architecture} for our implementation). Second, as the loss rate rises steeply toward low energies, a proton degraded to $E_{\rm K} \lesssim 1\,\mathrm{MeV}$ loses its remaining energy within a few percent of a Hubble time: protons injected at low energy by distant sources are absorbed in transit, which sets the low-energy end of the CR spectrum arriving at a given cell.

Only a fraction of the energy lost to excitation and ionization heats the gas, because the secondary electrons also excite and ionize. This fraction depends on the energy of the electrons that enter the secondary cascade. An X-ray photoelectron carries essentially all of the photon's energy, which is then partitioned with the fractions of \citet{FurlanettoStoever2010}; a CR proton instead imparts only a small part of its energy in each encounter, so the electrons it ejects are far less energetic. We therefore use the partition that \citet{SpitzerScott1969} computed for the secondaries of ionization by energetic particles in neutral hydrogen, as do GJ23, interpolated logarithmically in the free-electron fraction $x_{\rm e}$, so that $f_{\rm heat}$ rises from $0.10$ at $x_{\rm e} = 10^{-4}$ to $0.53$ at $x_{\rm e} = 0.5$, and is held constant outside that range.

In partially ionized gas, the proton also scatters off the thermal free electrons. The Coulomb loss rate, from eq.~(4.22) of \citet{Mannheim1994} (see also~\citep{Schlickeiser2002}), is
\begin{equation}
\left.\frac{\mathrm{d}E_{\rm K}}{\mathrm{d}t}\right|_{\rm C}
\simeq -3.10\times10^{-7}\,\mathrm{eV\,s^{-1}\,cm^{3}}\;
x_{\rm e}\, n_{\rm b}\,
\frac{\beta^{2}}{(T_{\rm K}/2.45\times10^{9}\,\mathrm{K})^{3/2} + \beta^{3}},
\label{eq:coulomb_loss}
\end{equation}
where $x_{\rm e}$ is the free-electron fraction, $n_{\rm b}$ the baryon number density, and $T_{\rm K}$ the local kinetic temperature, which we adopt for the thermal electrons. The thermal term plays the role of $\beta_0^{3}$ in eq.~\eqref{eq:bethe_bloch}, setting the velocity below which the proton is slower than the thermal electrons and the loss rate turns over. Unlike the energy lost to excitation and ionization, the energy lost to Coulomb scattering goes directly into the thermal electrons, so this channel heats the IGM with unit efficiency (section~\ref{sec:architecture}).

Combining the three channels, the kinetic energy of a CR proton in the (partially) neutral IGM evolves as
\begin{equation}
\frac{\mathrm{d}E_{\rm K}}{\mathrm{d}z}
 \,=\, \left.\frac{\mathrm{d}E_{\rm K}}{\mathrm{d}z}\right|_{\rm ad}
 + \left.\frac{\mathrm{d}E_{\rm K}}{\mathrm{d}t}\right|_{\rm E\&I}\frac{\mathrm{d}t}{\mathrm{d}z}
 + \left.\frac{\mathrm{d}E_{\rm K}}{\mathrm{d}t}\right|_{\rm C}\frac{\mathrm{d}t}{\mathrm{d}z} .
 \label{eq:energy_evolution}
\end{equation}
Since $\mathrm{d}t/\mathrm{d}z<0$, the (negative) collisional loss rates contribute positively to $\mathrm{d}E_{\rm K}/\mathrm{d}z$, i.e., the kinetic energy decreases as cosmic time advances. We integrate eq.~\eqref{eq:energy_evolution} numerically to obtain the energy $E_{\rm K}(E_{\rm K,i}; z, z'')$ at redshift $z$ of a CR proton emitted at redshift $z''$ with initial kinetic energy $E_{\rm K,i}$.

% ------------------------------------------------------------
\subsection{Momentum power-law injection spectrum}\label{sec:spectrum}

Supernova remnant shocks have long been identified as the principal accelerators of Galactic CRs \citep{BaadeZwicky1934, Vink2012, Blasi2013}. In DSA, particles gain energy by repeatedly crossing the shock front, which diffusive scattering lets them do many times \citep{Axford1977, Krymskii1977, Bell1978, BlandfordOstriker1978}. As both the fractional momentum gain per crossing and the escape probability are approximately independent of momentum, the accelerated particles follow a power law in momentum, not in kinetic energy \citep{Drury1983, MalkovDrury2001}. In test-particle DSA, the slope is set by the shock compression ratio, a strong non-relativistic shock giving a phase-space distribution $f(p)\propto p^{-4}$ \citep{Drury1983}; observed and simulated remnant spectra can be somewhat steeper, typically $\mathrm{d}N/\mathrm{d}E_{\rm K}\propto E_{\rm K}^{-2.2}$--$E_{\rm K}^{-2.4}$ at relativistic energies, e.g.~\citep{Vink2012,Caprioli2012,Morlino2012}. We therefore model the escaping proton spectrum as a momentum power law of free slope, as in~\citep{Leite2017, JanaNath2018}.

Here $\mathrm{d}N/\mathrm{d}p$ and $\mathrm{d}N/\mathrm{d}E_{\rm K}$ denote spectra per unit injected CR-proton energy. For an isotropic distribution, the number of CRs per unit momentum is
\begin{equation}
\frac{\mathrm{d}N}{\mathrm{d}p} = 4\pi N_0\, p^{2}\left(\frac{p}{m_{\rm p} c}\right)^{-\alpha},
\label{eq:dndp}
\end{equation}
where $p = |\mathbf{p}|$, $\alpha$ is the momentum spectral index, and $N_0$ is a normalization constant, fixed by requiring unit total kinetic energy,
\begin{equation}
1
= \int_{p_{\min}}^{p_{\max}} \mathrm{d}p\;\frac{\mathrm{d}N}{\mathrm{d}p}\,E_{\rm K}(p),
\label{eq:shape_norm}
\end{equation}
where
\begin{equation}
E_{\rm K}(p) = \sqrt{p^{2}c^{2} + m_{\rm p}^{2}c^{4}} - m_{\rm p} c^{2}.
\end{equation}
The cutoffs $p_{\min}$ and $p_{\max}$ correspond to kinetic energies $E_{\rm K,min}$ and $E_{\rm K,max}$ (section~\ref{sec:escape}).

We introduce the dimensionless momentum $\tilde{p} \equiv p/(m_{\rm p}c)$ and the integral
\begin{equation}
I(\alpha, \tilde{p}_{\min})
\equiv \int_{\tilde{p}_{\min}}^{\tilde{p}_{\max}} \mathrm{d}\tilde{p}\;
\tilde{p}^{\,2-\alpha}\left(\sqrt{1+\tilde{p}^{2}} - 1\right),
\label{eq:I_def}
\end{equation}
where $\tilde{p}_{\min} \equiv p_{\min}/(m_{\rm p} c)$ and $\tilde{p}_{\max} \equiv p_{\max}/(m_{\rm p} c)$ (we suppress the weak dependence of $I$ on $\tilde{p}_{\max}$), or equivalently
\begin{equation}
\tilde{p}_{\min}
=
\left[
\frac{E_{\rm K,min}}{m_{\rm p} c^2}
\left(
\frac{E_{\rm K,min}}{m_{\rm p} c^2}+2
\right)
\right]^{1/2}.
\end{equation}
The normalization condition then gives
\begin{equation}
N_0 = 1\,/\,\bigl[\,4\pi c\,(m_{\rm p} c)^{4}\,
I(\alpha, \tilde{p}_{\min})\,\bigr]\,.
\end{equation}
Substituting $N_0$ and changing variables to kinetic energy yields the injection spectrum we use,
\begin{equation}
\frac{\mathrm{d}N}{\mathrm{d}E_{\rm K}}
= \frac{(m_{\rm p} c^{2})^{\alpha-4}}
       {I(\alpha, \tilde{p}_{\min})}\,
  (E_{\rm K} + m_{\rm p} c^{2})\,
  \bigl[E_{\rm K}(E_{\rm K} + 2 m_{\rm p} c^{2})\bigr]^{(1-\alpha)/2}.
\label{eq:dndek}
\end{equation}
Multiplying eq.~\eqref{eq:dndek} by the CR energy released by one supernova, $\xi_{\rm cr}E_{\rm SN}$, gives the number spectrum of a single supernova.

Two limits clarify the shape of eq.~\eqref{eq:dndek}. In the non-relativistic regime, $E_{\rm K} \ll m_{\rm p} c^{2}$, it reduces to $\mathrm{d}N/\mathrm{d}E_{\rm K} \propto E_{\rm K}^{(1-\alpha)/2}$, while in the ultra-relativistic regime, $E_{\rm K} \gg m_{\rm p} c^{2}$, it steepens to $\mathrm{d}N/\mathrm{d}E_{\rm K} \propto E_{\rm K}^{2-\alpha}$. In this notation, test-particle DSA at a strong shock corresponds to $\alpha = 4$, and the observed relativistic slopes quoted above to $\alpha \approx 4.2$--$4.4$. It is the non-relativistic slope, however, that governs the sub-$40\,\mathrm{MeV}$ protons dominating IGM heating (section~\ref{sec:attribution}), and hence the heating efficiency of the model.

With no upper cutoff, the integral of eq.~\eqref{eq:I_def} converges only for $\alpha>4$, the regime we explore, so that $p_{\max}$ affects the normalization only weakly ($6\%$ for $\alpha = 4.2$ and $E_{\rm K,max} = 10^{9}\,\mathrm{MeV}$); for $\alpha \leq 4$, a finite maximum momentum would be required. At the low-momentum end, we impose a minimum escaping kinetic energy, $E_{\rm K,min}$, discussed in section~\ref{sec:escape}.

% ------------------------------------------------------------
\subsection{Injection efficiency and its expected range}\label{sec:escape}

The injection model thus has three free parameters: the spectral index $\alpha$, the low-energy cutoff $E_{\rm K,min}$, and an overall efficiency $\eta_{\rm cr}$. To source the unit-energy spectrum of eq.~\eqref{eq:dndek} from star-forming galaxies, we multiply it by the CR energy injected into the IGM per unit stellar mass formed,
\begin{equation}
\eta_{\rm cr} \equiv f_{\rm esc}\,f_{\rm SN}\,\xi_{\rm cr}\,E_{\rm SN} ,
\label{eq:eta_cr_def}
\end{equation}
in units of $\mathrm{erg}\,M_\odot^{-1}$; it is the product of four factors that differ markedly in how well they are constrained: the fraction $f_{\rm esc}$ of the accelerated CR-proton energy that escapes into the IGM; the number of core-collapse supernovae per unit stellar mass formed,
\begin{equation}
f_{\rm SN} = \frac{\int_{m_{\rm SN}}^{m_{\rm BH}} \phi(m)\,\mathrm{d}m}
{\int_{m_{\rm low}}^{m_{\rm up}} m\,\phi(m)\,\mathrm{d}m},
\label{eq:fsn}
\end{equation}
where $\phi(m)$ is the stellar initial mass function (IMF),
$m_{\rm SN} \approx 8\,M_\odot$ is the minimum progenitor mass for core
collapse, $m_{\rm BH} \approx 40\,M_\odot$ is the mass above which stars are expected to collapse directly to black holes without a successful explosion \citep{Heger2003}, and $m_{\rm low}$ and $m_{\rm up}$ are the limits of the IMF itself, over which the stellar mass in the denominator is integrated.

A Salpeter IMF~\citep{Salpeter1955} over $0.1$--$100\,M_\odot$ gives $f_{\rm SN} \approx 7 \times 10^{-3}\,M_\odot^{-1}$, the value adopted by \citet{MadauDickinson2014}, and the Kroupa~\citep{Kroupa2001} and Chabrier~\citep{Chabrier2003} forms give $\approx (1.0$--$1.1) \times 10^{-2}\,M_\odot^{-1}$, i.e., roughly one core-collapse supernova per $100\,M_\odot$ of stars formed; such values reproduce the observed core-collapse rate to within a factor of $\approx 2$~\citep{Horiuchi2011, MadauDickinson2014}. Each such supernova releases $E_{\rm SN} \sim 10^{51}\,\mathrm{erg}$. The gamma-ray emission of Galactic supernova remnants and the Galactic CR energy budget together indicate $\xi_{\rm cr} \sim 0.05$--$0.3$, e.g.~\citep{Vink2012, Caprioli2012, Morlino2012}, with non-linear shock modification expected to limit it to a few tens of percent \citep{MalkovDrury2001}; as it is poorly determined for the first galaxies, we treat it as free. The escape fraction is the least constrained factor at high redshift, depending on the magnetic-field geometry and on the gas column traversed before the CRs are confined or lose their energy.

Together, these factors cap the efficiency that the CR channel can deliver. With $E_{\rm SN} \approx 10^{51}\,\mathrm{erg}$ and $f_{\rm SN} \approx 10^{-2}\,M_\odot^{-1}$, eq.~\eqref{eq:eta_cr_def} gives $\eta_{\rm cr} \approx 10^{49}\,f_{\rm esc}\,\xi_{\rm cr}\, \mathrm{erg}\,M_\odot^{-1}$, so that the observed range of $\xi_{\rm cr}$ with full escape corresponds to $\eta_{\rm cr} \sim 5 \times 10^{47}$--$3 \times 10^{48}\,\mathrm{erg}\,M_\odot^{-1}$. Our fiducial value, $10^{48}\,\mathrm{erg}\,M_\odot^{-1}$, lies within this range for $f_{\rm esc}\xi_{\rm cr} \approx 0.1$, whereas $10^{49}\,\mathrm{erg}\,M_\odot^{-1}$ lies above it, and is not attainable by a Pop~II population for any admissible combination of escape and acceleration efficiencies.

The ceiling is higher for Pop~III stars, whose characteristic mass is expected to be higher in the absence of metal cooling: although no observation constrains their IMF, hydrodynamic simulations of primordial star formation suggest a top-heavy distribution extending to several hundred solar masses~\citep{BrommLarson2004, Hirano2014}, though disk fragmentation may lower the typical mass~\citep{Grief11}. For illustrative top-heavy forms, eq.~\eqref{eq:fsn} gives $f_{\rm SN} \approx 1.5 \times 10^{-2}\,M_\odot^{-1}$ for a log-flat IMF over $10$--$100\,M_\odot$, and $4.0 \times 10^{-2}\,M_\odot^{-1}$ for a Salpeter slope over the same range. Stars of $140$--$260\,M_\odot$ are moreover predicted to explode as pair-instability supernovae, the pair instability reopening explodability above the direct-collapse range, releasing $\sim 3\times10^{51}$--$10^{53}\, \mathrm{erg}$ \citep{HegerWoosley2002, Heger2003}. Even if all Pop~III stellar mass formed in the most massive of these ($\sim 260\,M_\odot$ stars, each releasing $\sim 10^{53}\,\mathrm{erg}$), with $\xi_{\rm cr} = 0.3$ and full escape, eq.~\eqref{eq:eta_cr_def} gives $\eta_{\rm cr} \lesssim 0.3 \times 10^{53}\,\mathrm{erg} / 260\,M_\odot \approx 10^{50}\,\mathrm{erg}\,M_\odot^{-1}$. As these estimates rest on assumed IMFs, our optimistic model ($\eta_{\rm cr} = 10^{49}\,\mathrm{erg}\,M_\odot^{-1}$; section~\ref{sec:sims}) should be read as a limiting case, attainable only if early star formation was dominated by massive Pop~III progenitors whose CRs escaped efficiently into the IGM. We therefore vary $\eta_{\rm cr}$, the single parameter through which the CR injection couples to the galaxy model, over orders of magnitude.

The low-energy cutoff is physical rather than mathematical. The injection of particles from the thermal pool into DSA is not captured by the power-law solution, and the lowest-energy protons might lose their energy in the host ISM before reaching the diffuse IGM \citep{JanaNath2018}; for our indices the normalization integral converges as $p_{\min} \to 0$ (it diverges only for $\alpha \geq 5$), although the number of injected protons diverges in that limit for $\alpha \geq 3$. Lowering $E_{\rm K,min}$ enhances prompt heating close to the sources, while raising it makes the heating field smoother and more delayed. We adopt $E_{\rm K,min} = 0.01\,\mathrm{MeV}$ and $E_{\rm K,max} = 10^{9}\,\mathrm{MeV}$, as in the reference model of GJ23. The lower cutoff must be quoted alongside any comparison of injection spectra, since the energy fractions of section~\ref{sec:attribution} depend on where it is placed: the kinetic-energy power law carries equal energy per decade, so its fraction below $30\,\mathrm{MeV}$ grows logarithmically as $E_{\rm K,min}$ is lowered.

With the unit-energy spectrum of eq.~\eqref{eq:dndek} and the efficiency $\eta_{\rm cr}$, common to all star-forming halos (but see section~\ref{sec:direct_injection_physics}), the CR injection is fully specified for a given star formation rate. In the next section, we couple it to the galaxy populations of \texttt{21cmFAST}, and propagate the injected protons through the losses of section~\ref{sec:losses} to obtain the heating rate in each cell.

% ============================================================
\section{Implementation in \texttt{21cmFAST}}\label{sec:implementation}

% ------------------------------------------------------------
% ------------------------------------------------------------

\subsection{Galaxy populations}\label{sec:galaxies}

%The cosmic ray injection rate constructed in Section~\ref{sec:heating_rate}
%(Eq.~\ref{eq:cr_injection}) is sourced by the star formation rate density of the
%host galaxies, so 
Star formation proceeds in two populations~\citep{Qin20, Munoz2022}: Pop~II stars in ACGs, hosted by halos above the atomic-cooling threshold, $M_{\rm atom}$, and Pop~III stars in MCGs, hosted by less massive halos, in which the gas cools through molecular hydrogen. We summarize here only the features of the model that enter our CR calculation, and refer the reader to \citet{Munoz2022} for a complete description.

The galaxy model enters the CR calculation through the star formation rate density (SFRD) of each population, obtained by integrating the SFR per halo over the conditional halo mass function. For ACGs, the SFRD is
\begin{equation}
\label{eq:sfrd_acg}
\mathrm{SFRD}_{\rm II}(\mathbf{x}, z) =
\int \mathrm{d}M_{\rm h}\, \frac{\mathrm{d}n}{\mathrm{d}M_{\rm h}}\,
\dot{M}_\star^{(\rm II)}\, f_{\rm duty}^{(\rm II)},
\end{equation}
where $\mathrm{d}n/\mathrm{d}M_{\rm h}$ is the conditional Press--Schechter mass function~\citep{LC93} evaluated on the evolved non-linear density field (section~\ref{sec:architecture}), $\dot{M}_\star^{(\rm II)}$ is the Pop~II SFR per halo, and the duty cycle $f_{\rm duty}^{(\rm II)} = \exp(-M_{\rm turn}^{(\rm II)}/M_{\rm h})$ suppresses star formation in halos below the turnover mass $M_{\rm turn}^{(\rm II)} = \max(M_{\rm ion,crit}, M_{\rm atom})$ (see below). The Pop~III SFRD takes the same form,
\begin{equation}
\label{eq:sfrd_mcg}
\mathrm{SFRD}_{\rm III}(\mathbf{x}, z) =
\int \mathrm{d}M_{\rm h}\, \frac{\mathrm{d}n}{\mathrm{d}M_{\rm h}}\,
\dot{M}_\star^{(\rm III)}\, f_{\rm duty}^{(\rm III)},
\end{equation}
but with a duty cycle $f_{\rm duty}^{(\rm III)} = \exp(-M_{\rm mol,crit}/M_{\rm h})\,\exp(-M_{\rm h}/M_{\rm atom})$, whose two factors confine MCG star formation to a mass window. The first imposes a lower limit at the molecular-cooling threshold, $M_{\rm mol,crit}(\mathbf{x}, z)$, below which halos are too small to cool through molecular hydrogen and form Pop~III stars. The second hands star formation over to Pop~II above $M_{\rm atom}$, where halos cool atomically, as we assume that ACGs have been pre-enriched by their MCG progenitors.

Both SFRs follow from a population-averaged stellar-to-halo mass relation (SHMR), which we take to be a power law in halo mass,
\begin{equation}
\label{eq:shmr}
M_\star^{(i)} = f_\star^{(i)}\, \frac{\Omega_{\rm b}}{\Omega_{\rm m}}
\left(\frac{M_{\rm h}}{M_{(i)}}\right)^{\alpha_\star^{(i)}} M_{\rm h},
\end{equation}
where $i \in \{\mathrm{II}, \mathrm{III}\}$ labels the population, $\Omega_{\rm b}/\Omega_{\rm m}$ is the cosmic baryon fraction, $f_\star^{(i)}$ is the stellar fraction at a pivot halo mass $M_{(i)}$ ($10^{10}\,M_\odot$ for ACGs and $10^{7}\,M_\odot$ for MCGs), and $\alpha_\star^{(i)}$ is the power-law index. The SFR per halo is $\dot{M}_\star^{(i)} = M_\star^{(i)} / (t_\star H^{-1}(z))$, where $t_\star = 0.5$ is the star-formation timescale in units of the Hubble time, $H^{-1}(z)$. The ACG normalization and index, $f_\star^{(\rm II)} = 10^{-1.25}$ and $\alpha_\star^{(\rm II)} = 0.5$, are calibrated to observed UV luminosity functions \citep{Park2019, Bouwens2021}. The MCG parameters cannot be constrained by these data, and we fix them to the EOS21 values, $f_\star^{(\rm III)} = 10^{-2.5}$ and $\alpha_\star^{(\rm III)} = 0$, i.e., a stellar fraction independent of halo mass (for priors on MCG star formation informed by hydrodynamic simulations, with $\alpha_\star^{(\rm III)} = \alpha_\star^{(\rm II)}$, see~\citep{Breitman2026}).

The positive index $\alpha_\star^{(\rm II)}$ reflects stellar feedback: as supernova-driven outflows remove a larger fraction of the gas supply from shallower potential wells, the stellar fraction rises with halo mass, with $\alpha_\star \approx 2/3$ ($1/3$) predicted for energy-driven (momentum-driven) outflows \citep{Furlanetto2017}, bracketing the UV-calibrated value $\alpha_\star^{(\rm II)} = 0.5$ \citep{Park2019}. This is a key difference from the constant stellar-to-halo mass ratio ($\alpha_\star = 0$) adopted by GJ23: by suppressing star formation in the lowest-mass halos, it results in a more rapid evolution of the SFRD and less star formation at high redshift, and hence less CR injection (eq.~\eqref{eq:cr_shell_integration}); we quantify this effect in section~\ref{sec:attribution}.
%; we quantify its contribution to

The turnover masses are set by feedback. For both populations, photoheating during reionization raises the minimum halo mass for star formation, as gas in halos exposed to a strong ionizing background is heated above their virial temperature; we capture this through a critical mass, $M_{\rm ion,crit}$, that depends on the local ionizing background and on the redshift at which the cell was reionized, using the functional form of \citet{SobacchiMesinger2013}. For ACGs this is the only additional feedback. MCGs, residing in shallower potential wells, are also sensitive to the photodissociation of molecular hydrogen by LW photons \citep{Machacek2001} and to the suppression of gas accretion by the dark matter--baryon relative (streaming) velocity \citep{TseliakhovichHirata2010}, both of which raise $M_{\rm mol,crit}$. We model their joint effect as
\begin{equation}
\label{eq:mmol}
M_{\rm mol,crit}(\mathbf{x}, z) = M_0(z)\, f_{v_{\rm cb}}(v_{\rm cb})\, f_{\rm LW}(J_{21}),
\end{equation}
where $M_0(z) \propto (1+z)^{-3/2}$ is the no-feedback threshold, corresponding to a virial temperature $T_{\rm vir} \approx 10^{3}\,\mathrm{K}$, and $f_{v_{\rm cb}}$ and $f_{\rm LW}$ are dimensionless factors, each equal to unity in the absence of the corresponding feedback. The factor $f_{v_{\rm cb}}$ increases with the local streaming velocity $v_{\rm cb}$, following the calibration of \citet{Munoz2022}, and
\begin{equation}
\label{eq:flw}
f_{\rm LW}(J_{21}) = 1 + A_{\rm LW}\, J_{21}^{\,\beta_{\rm LW}},
\end{equation}
where $J_{21}$ is the spatially varying LW intensity in units of $10^{-21}\,\mathrm{erg\,s^{-1}\,cm^{-2}\,Hz^{-1}\,sr^{-1}}$, and $A_{\rm LW} = 2$ and $\beta_{\rm LW} = 0.6$ are fitting parameters. These forms are calibrated against the hydrodynamic simulations of \citet{Schauer2021} and \citet{Kulkarni2021}, which resolve LW self-shielding and streaming-velocity feedback in the host halos of the first MCGs (see \citet{Munoz2022} for details of the fit).

\begin{figure}
\centering
\includegraphics[width=\columnwidth]{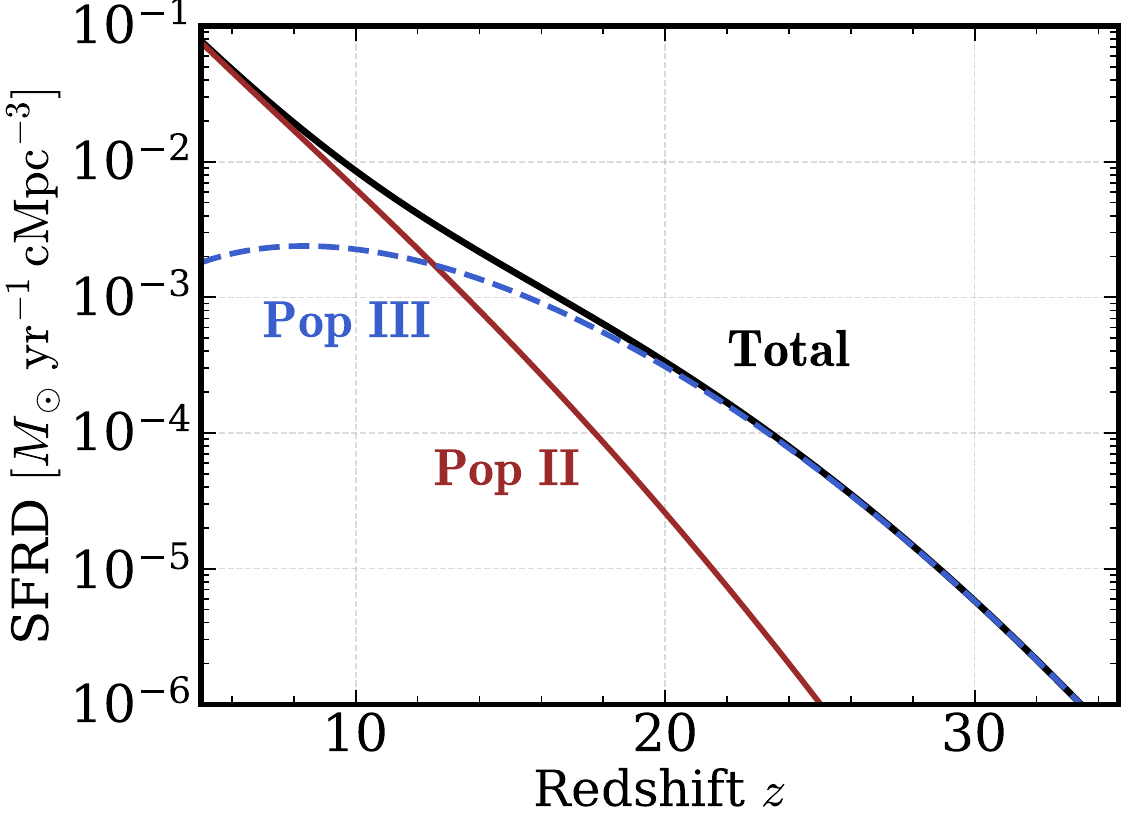}
\caption{Volume-averaged SFRD of Pop~II stars in ACGs (solid red curve), Pop~III stars in MCGs (dashed blue curve), and their sum (solid black curve), for the fiducial EOS21 parameters. The CR injection of eq.~\eqref{eq:cr_shell_integration} is sourced by the sum.}
\label{fig:sfrd}
\end{figure}

The LW intensity is evolved inhomogeneously on the simulation grid, sourced by the SFRDs of eqs.~\eqref{eq:sfrd_acg} and~\eqref{eq:sfrd_mcg}; as star formation builds up, $J_{21}$, and hence $M_{\rm mol,crit}$, rise monotonically through cosmic dawn. As the halo mass function also shifts toward more massive halos, the transition from MCG- to ACG-dominated star formation would occur even without feedback, but LW feedback brings it forward to higher redshift. In figure~\ref{fig:sfrd} we show the resulting SFRDs of the two populations for our fiducial parameters: MCGs dominate star formation at $z \gtrsim 12.5$, and ACGs thereafter.

\subsection{Cosmic-ray transport and heating rate}\label{sec:architecture}

We implement CR heating in a modified version of \texttt{21cmFAST} v3~\citep{Murray2020}\footnote{\url{https://github.com/michaelpaganocosmo/21cmFAST-CosmicRays}}, alongside the existing X-ray, Ly$\alpha$ and Compton heating channels. Outside fully ionized H\,\textsc{ii} regions, the kinetic temperature $T_{\rm K}$ and free-electron fraction $x_{\rm e}$ of the gas are evolved in every cell by integrating coupled equations in redshift, to which CR heating adds one channel, with a heating rate per baryon, $\varepsilon_{\rm CR}$, computed with the same source infrastructure as X-ray heating \citep{Mesinger2011, Mesinger2013}.

At each redshift step, $z'$, the contribution of luminous sources is accumulated over concentric shells around each cell, each corresponding to an earlier redshift, $z'' > z'$, along the past light cone, with the emissivity of each shell set by the SFRDs of section~\ref{sec:galaxies}. The conditional Press--Schechter mass function entering these SFRDs is evaluated on the evolved non-linear density field smoothed on the shell radius, $R''$, and normalized so that its box-averaged mean reproduces the Sheth--Tormen mass function~\citep{ST99}, as in~\citep{Mesinger2011, Mesinger2013}. X-rays emitted in each shell are transported to the cell as an angle-averaged specific intensity attenuated by the spatially averaged IGM opacity, and the absorbed energy is partitioned into heat, secondary ionizations and Ly$\alpha$ photons with the energy-dependent fractions of \citet{FurlanettoStoever2010}.

Our CR implementation uses the same infrastructure, but injects protons rather than photons. The CR emissivity is sourced by the total SFRD, $\mathrm{SFRD}_{\rm tot} = \mathrm{SFRD}_{\rm II} + \mathrm{SFRD}_{\rm III}$, i.e., the sum of the ACG and MCG contributions of eqs.~\eqref{eq:sfrd_acg} and~\eqref{eq:sfrd_mcg}, each with its own SHMR but with a common efficiency $\eta_{\rm cr}$ (the population-dependent case is treated in section~\ref{sec:direct_injection_physics}). A galaxy forming stars at redshift $z''$ injects CRs into the surrounding IGM with the momentum power-law spectrum of eq.~\eqref{eq:dndek} and the efficiency $\eta_{\rm cr}$ of eq.~\eqref{eq:eta_cr_def}. The proper number density of CR protons present at the cell at $z'$, $n_{\rm CR}$, per unit initial kinetic energy and per unit source redshift, is then
\begin{equation}
\label{eq:cr_shell_integration}
\frac{\mathrm{d}^2 n_{\rm CR}}{\mathrm{d}E_{\rm K,i}\,\mathrm{d}z''}(\mathbf{x}, z')
= (1+z')^3\,\eta_{\rm cr}\,\overline{\mathrm{SFRD}}_{\rm tot}(\mathbf{x}, z'')\,
\left|\frac{\mathrm{d}t}{\mathrm{d}z''}\right|\,
\frac{\mathrm{d}N}{\mathrm{d}E_{\rm K}}\bigg|_{E_{\rm K,i}} ,
\end{equation}
where $E_{\rm K,i}$ is the kinetic energy at injection, $\overline{\mathrm{SFRD}}_{\rm tot}(\mathbf{x}, z'')$ is the total SFRD at $z''$ evaluated on the density field smoothed on the shell radius $R''$ around $\mathbf{x}$, $|\mathrm{d}t/\mathrm{d}z''|$ converts the emission rate into the number of protons emitted per unit source redshift, the factor $(1+z')^3$ converts comoving into proper density, and $\mathrm{d}N/\mathrm{d}E_{\rm K}$ is the unit-energy spectrum of eq.~\eqref{eq:dndek}.

Unlike X-ray photons, which either are absorbed or arrive with their emitted energy, a CR proton is slowed continuously: between its injection at $z''$ and its arrival at the cell at $z'$, it loses energy to excitation and ionization, Coulomb scattering and adiabatic expansion, so that the arriving spectrum is shifted in energy rather than reduced in normalization. We integrate eq.~\eqref{eq:energy_evolution} along the past light cone to obtain the arrival energy $E_{\rm K}(E_{\rm K,i}; z', z'')$ of a proton injected at $z''$ with energy $E_{\rm K,i}$; protons injected at low energy at sufficiently high $z''$ lose all of their energy in transit. As the arrival energy depends on the source redshift, we evaluate it shell by shell, as \texttt{21cmFAST} does for the X-ray optical depth, $\tau_{\rm X}$.

Assigning each shell its emission redshift through the light-travel time, as for X-rays, implies that the protons reach the cell at speed $c$. Sub-relativistic protons instead travel at $\beta c$, and diffusively if the IGM is magnetized \citep{Leite2017, JanaNath2018}, so this approximation spreads the heat of each proton over a region larger than it can physically reach by a factor of order $\beta^{-1}$: at $z = 10$ a $1\,\mathrm{MeV}$ proton ($\beta \approx 0.05$) thermalizes in $\approx 20\,\mathrm{Myr}$, during which light travels $\approx 70\,\mathrm{cMpc}$ but the proton itself only $\approx 2\,\mathrm{cMpc}$. What the shell integration retains is therefore not the proton's range but its lifetime: protons that thermalize before the light-travel time to the cell contribute nothing, so the CR heating field still has a shorter effective horizon than the X-ray one, set by the thermalization time rather than by the stopping distance (middle panel of figure~\ref{fig:sed}). Our CR heating field is consequently smoother than the physical one, and the scale dependence of the differences from X-ray heating that we find in section~\ref{sec:power_spectrum} is likely a lower bound on the true signature.

The local CR heating rate per baryon follows by integrating, over source redshift and injection efficiencies, the rate at which the arriving protons heat the gas,
\begin{equation}
\label{eq:cr_spectral_integration}
\varepsilon_{\rm CR}(\mathbf{x}, z') = \frac{1}{n_{\rm b}(\mathbf{x}, z')}
\int_{z'}^{z''_{\max}} \mathrm{d}z'' \int_{E_{\rm K,min}}^{E_{\rm K,max}} \mathrm{d}E_{\rm K,i}\,
\frac{\mathrm{d}^2 n_{\rm CR}}{\mathrm{d}E_{\rm K,i}\,\mathrm{d}z''}\,
\mathcal{L}\big[E_{\rm K}(E_{\rm K,i}; z', z'')\big] ,
\end{equation}
where $n_{\rm b}(\mathbf{x}, z')$ is the local proper baryon number density, $z''_{\max}$ is the redshift of the outermost shell, at a comoving radius $R_{\max} = 500\,\mathrm{cMpc}$, as for X-ray heating, and
\begin{equation}
\mathcal{L}(E_{\rm K}) \equiv f_{\rm heat}(x_{\rm e}) \left|\frac{\mathrm{d}E_{\rm K}}{\mathrm{d}t}\right|_{\rm E\&I}
+ \left|\frac{\mathrm{d}E_{\rm K}}{\mathrm{d}t}\right|_{\rm C}
\end{equation}
is the rate at which a proton of arrival energy $E_{\rm K}$ heats the gas, with the heat fraction $f_{\rm heat}(x_{\rm e})$ of section~\ref{sec:losses} and the loss rates of eqs.~\eqref{eq:bethe_bloch} and~\eqref{eq:coulomb_loss}, evaluated with the local $x_{\rm e}$ and $T_{\rm K}$; $\mathcal{L}=0$ for protons that thermalized before reaching the cell.

% ---------------------------------------------------------------------------
%  NOTE (f_heat for CR secondaries is conservative) -- not for print.
%
% If a questions arises whether the Spitzer & Scott (1969) partition
% underestimates the heat from CR secondaries, independent support:
%
%   arXiv:astro-ph/9311048, "Cosmic ray ionization of the interstellar
%   medium", sec. (e). Taking the mean secondary-electron energy from CR
%   ionization as ~36 eV (Dalgarno & Griffing 1958), and following the
%   prescription of Spitzer & Tomasko (1968, ApJ 152, 971), they find a heat
%   source of 6.3 eV per primary ionizing event, i.e. a heating efficiency of
%   0.175 -- above the 0.10 our interpolated table assigns at x_e = 1e-4.
%
% Caveats before citing:
%   - authors, journal and x_e of that calculation not yet confirmed (ISM,
%     "weakly ionized"; check whether its x_e is comparable to ours);
%   - their cascade stops at 10.2 eV (excitation threshold), not 13.6 eV;
%   - Dalgarno & Griffing (1958) reference details unverified.
%
% Physical point it supports: secondaries ejected near threshold have little
% headroom for further ionization, so f_heat for CR secondaries should exceed
% the asymptotic value tabulated for energetic electrons.
% ---------------------------------------------------------------------------

The resulting rate $\varepsilon_{\rm CR}$ is added to the sum over heating channels, $\sum_j \varepsilon_j$, and the kinetic temperature is advanced to the next redshift step through
\begin{equation}
\label{eq:tk_temperature_update}
\frac{\mathrm{d}T_{\rm K}(\mathbf{x}, z')}{\mathrm{d}z'} =
\frac{2}{3k_{\rm B}\,[1 + x_{\rm e}(\mathbf{x}, z')]}\,
\frac{\mathrm{d}t}{\mathrm{d}z'} \sum_{j} \varepsilon_{j}
+ \frac{2}{3} \frac{T_{\rm K}}{n_{\rm b}} \frac{\mathrm{d}n_{\rm b}}{\mathrm{d}z'}
- \frac{T_{\rm K}}{1 + x_{\rm e}} \frac{\mathrm{d}x_{\rm e}}{\mathrm{d}z'} ,
\end{equation}
where the first term collects the heating rates per baryon, $\varepsilon_j$, of all channels $j$, the second describes adiabatic cooling (and heating) as the local baryon density evolves, which for the mean density reduces to $2T_{\rm K}/(1+z')$, and the third the change in the number of particles due to ionization. We carry out this calculation in every cell of the simulation, so that the CR heating field inherits the full spatial structure of the source distribution.

\subsection{MCG-only injection}\label{sec:direct_injection_physics}

So far, CRs escape all star-forming halos through the same diffusive process, parametrized by $\eta_{\rm cr}$. This is a reasonable description for ACGs, in which the supernova remnant forms inside a dense ISM and CRs leak out over tens of Myr \citep{Schlickeiser2002, Leite2017}. In the lowest-mass Pop~III hosts, however, CRs could escape differently. \citet{SazonovSunyaev2015} argued that, in halos below $\sim 10^{7}\,M_\odot$, the Pop~III progenitor photoevaporates the gas out to (and beyond) the virial radius before exploding, so that the remnant expands into an essentially evacuated region, forms a strong shock only once it overtakes the photoionized gas at several virial radii, and dissipates only $1$--$10\,\mathrm{Myr}$ later. As the recombination time at these densities is shorter than this dissipation time, the surrounding gas has recombined by then, and the CRs are deposited directly into the neutral IGM, bypassing the ISM of the host. We refer to this mechanism as \emph{MCG-only} CR injection, as its distinguishing feature is the restriction of CR sources to MCGs below a maximum halo mass.

This picture, based on numerical results for photoevaporation, remnant propagation and recombination in low-mass Pop~III hosts \citep{KitayamaYoshida2005, Greif2007, Whalen2008}, applies only below a maximum halo mass, $M_{\rm cr}^{\rm max}$, conventionally $10^{7}\,M_\odot$, above which the progenitor cannot photoevaporate the gas out to the virial radius. Halos below $M_{\rm cr}^{\rm max}$ host only Pop~III stars (for $M_{\rm cr}^{\rm max} \leq 10^{7}\,M_\odot$, which lies below the atomic-cooling mass at $z \lesssim 20$), as Pop~II stars form in ACGs, which lie above the atomic-cooling threshold and have been enriched by a previous Pop~III generation (section~\ref{sec:galaxies}); CR injection is then dominated by MCGs. ACGs still accelerate CRs, but these must traverse the dense ISM of their hosts (section~\ref{sec:escape}), so their escape fraction is expected to be much smaller. We take the limiting case $f_{\rm esc}^{\rm ACG} \ll f_{\rm esc}^{\rm MCG}$ and set the ACG contribution to zero, bracketing the scenario from one side; the unrestricted injection of section~\ref{sec:escape}, with a common $\eta_{\rm cr}$, brackets it from the other.

We treat $M_{\rm cr}^{\rm max}$ as a free parameter rather than fixing it at $10^{7}\,M_\odot$, both to accommodate the uncertainty in the photoevaporation argument and because, as we show in section~\ref{sec:direct_injection_results}, the results depend sensitively on its value. The CR-emitting SFRD entering eq.~\eqref{eq:cr_shell_integration} is then
\begin{equation}
{\rm SFRD}_{\rm cr}(\mathbf{x}, z) =
\int^{M_{\rm cr}^{\rm max}} \mathrm{d}M_{\rm h}\,
\frac{\mathrm{d}n}{\mathrm{d}M_{\rm h}}\, \dot{M}_\star^{(\rm III)}\,
f_{\rm duty}^{(\rm III)} ,
\label{eq:sfrd_direct_injection}
\end{equation}
that is, the Pop~III SFRD of eq.~\eqref{eq:sfrd_mcg} truncated sharply at
$M_{\rm cr}^{\rm max}$. The lower end of the emitting window is not sharp:
it is set by the factor $\exp(-M_{\rm mol,crit}/M_{\rm h})$ in
$f_{\rm duty}^{(\rm III)}$, where $M_{\rm mol,crit}(\mathbf{x}, z)$ is the
redshift- and environment-dependent molecular-cooling threshold of
eq.~\eqref{eq:mmol}, set by the LW feedback prescription of
\citet{Munoz2022} and \citet{Kulkarni2021}. This replaces the sum of the MCG and ACG contributions at all halo masses.

Combining a fixed upper cutoff with a rising lower threshold leads to a qualitatively new behavior. As the LW background grows, $M_{\rm mol,crit}$ rises monotonically (section~\ref{sec:galaxies}); as it approaches and then exceeds $M_{\rm cr}^{\rm max}$, the duty cycle
suppresses star formation throughout the emitting window, and CR injection
declines exponentially, falling to $e^{-1}$ of its unsuppressed value at the top of the window when $M_{\rm mol,crit} = M_{\rm cr}^{\rm max}$ and to a few percent once $M_{\rm mol,crit} \approx 3\,M_{\rm cr}^{\rm max}$, even though Pop~III star formation continues in halos above $M_{\rm cr}^{\rm max}$, which by construction inject no CRs in this scenario. The redshift of this shutoff is set by $M_{\rm cr}^{\rm max}$ and by the LW history; we study its consequences in section~\ref{sec:direct_injection_results}.

\subsection{21-cm observables}\label{sec:21cm}

From the evolved kinetic and spin temperatures, together with the ionization and density fields, \texttt{21cmFAST} computes the 21-cm brightness temperature in every cell. Along a line of sight at observed frequency $\nu$, radiative transfer through the line gives the differential brightness temperature against the CMB \citep{MadauMeiksinRees1997, Furlanetto2006review, Mesinger2011},
\begin{equation}
\delta T_b(\nu) = \frac{T_{\rm S} - T_{\rm CMB}(z)}{1+z}
\left( 1 - e^{-\tau_{\nu_0}} \right),
\label{eq:dTb_exact}
\end{equation}
where $T_{\rm S}$ is the spin temperature, $T_{\rm CMB}(z)$ is the CMB temperature, and $\tau_{\nu_0}$ is the optical depth at line center,
\begin{equation}
\tau_{\nu_0} = \frac{3 c^3 \hbar A_{10} \, n_{\rm HI}}
{16 k_{\rm B} \nu_{21}^2 \, T_{\rm S} \left[ H(z) + \mathrm{d}v_r/\mathrm{d}r \right]},
\label{eq:tau}
\end{equation}
in which $A_{10}$ is the spontaneous emission coefficient of the transition, $n_{\rm HI}$ is the proper number density of neutral hydrogen, $\nu_{21} = 1420\,\mathrm{MHz}$ is the rest-frame frequency of the transition (so that $\nu = \nu_{21}/(1+z)$), $H(z)$ is the Hubble rate, and $\mathrm{d}v_r/\mathrm{d}r$ is the line-of-sight gradient of the peculiar velocity. As $\tau_{\nu_0} \sim 0.01$--$0.1$ through most of cosmic dawn, eq.~\eqref{eq:dTb_exact} can be expanded to first order in $\tau_{\nu_0}$, giving the form we use throughout,
\begin{equation}
\label{eq:dTb}
\begin{split}
\delta T_b(\nu) \approx{}& 27\, x_{\rm HI}\,(1 + \delta_{\rm nl})
\left(\frac{H}{\mathrm{d}v_r/\mathrm{d}r + H}\right)
\left(1 - \frac{T_{\rm CMB}}{T_{\rm S}}\right) \\
&\times \left(\frac{1+z}{10}\,\frac{0.15}{\Omega_{\rm m} h^2}\right)^{1/2}
\left(\frac{\Omega_{\rm b} h^2}{0.023}\right)\,\mathrm{mK},
\end{split}
\end{equation}
where $x_{\rm HI}$ is the neutral hydrogen fraction, $\delta_{\rm nl}$ is the evolved non-linear density contrast, and the last two factors carry the cosmological dependence. The sign and amplitude of $\delta T_b$ are set by the factor $(1 - T_{\rm CMB}/T_{\rm S})$. Once Wouthuysen--Field coupling \citep{Wouthuysen1952, Field1958} by Ly$\alpha$ photons drives $T_{\rm S}$ toward the kinetic temperature $T_{\rm K}$, $\delta T_b$ traces $T_{\rm K}$ relative to $T_{\rm CMB}$.

We characterize the signal through its sky average, the global signal,
\begin{equation}
\label{eq:global_signal}
\langle \delta T_b \rangle(z) \equiv \frac{1}{V} \int_V \delta T_b(\mathbf{x}, z)\,
\mathrm{d}V,
\end{equation}
where the angle brackets denote the average over the simulation volume $V$, and through its fluctuations, quantified by the spherically averaged power spectrum,
\begin{equation}
\label{eq:power_spectrum}
\begin{aligned}
\langle \widetilde{\delta T_b}(\mathbf{k}, z)\,
\widetilde{\delta T_b}^{*}(\mathbf{k}', z) \rangle
&= (2\pi)^3\, \delta_{\rm D}(\mathbf{k} - \mathbf{k}')\, P(k, z), \\
\Delta^2(k, z) &\equiv \frac{k^3}{2\pi^2}\, P(k, z),
\end{aligned}
\end{equation}
where $\widetilde{\delta T_b}(\mathbf{k}, z)$ is the Fourier transform of the brightness temperature field, $\delta_{\rm D}$ is the Dirac delta function, $P(k, z)$ is the power spectrum, and the angle brackets denote an ensemble average, which we estimate by averaging over spherical shells of fixed $k = |\mathbf{k}|$. The dimensionless power spectrum, $\Delta^2(k, z)$, in units of $\mathrm{mK}^2$, is the contribution to the variance of $\delta T_b$ per logarithmic interval in $k$, and is the central statistic of section~\ref{sec:power_spectrum}.

\subsection{Simulations and reference models}
\label{sec:sims}

All results in section~\ref{sec:Results} are drawn from a fixed suite of simulations run with our modified \texttt{21cmFAST}. All runs share the same initial conditions, so that differences between models isolate the heating channel. Unless stated otherwise, every run adopts the EOS21 astrophysical parameters of \citet{Munoz2022}, including MCGs and Ly$\alpha$~\citep{Chuzhoy2007, Reis2021} and CMB heating~\citep{Venumadhav2018}, which the public \texttt{21cmFAST} implements following \citet{Reis2021} and \citet{Meiksin2021}, respectively; the CR models include no X-ray heating, and the X-ray-only reference model, with the EOS21 soft-band ($<2\,\mathrm{keV}$) luminosity $L_{\rm X}/\mathrm{SFR} = 10^{40.5}\,\mathrm{erg\,s^{-1}}\,M_\odot^{-1}\,\mathrm{yr}$, includes no CR heating. In section~\ref{sec:global_signal_eos21}, we disable Ly$\alpha$ and CMB heating to isolate the direct thermal effect of the CR and X-ray channels. We show the volume-averaged kinetic temperature in sections~\ref{sec:tk_evolution} and~\ref{sec:attribution}, and its average over the neutral IGM, $\langle T_{\rm K}\rangle_{\rm HI}$, in sections~\ref{sec:global_signal_eos21} and~\ref{sec:direct_injection_results}, since ionized regions in \texttt{21cmFAST} carry a separate temperature field.

We use two grid configurations. The heating rates, thermal histories and global signals of
sections~\ref{sec:heating_rate_per_energy_bin}, \ref{sec:tk_evolution}, \ref{sec:attribution},
\ref{sec:global_signal_eos21} and~\ref{sec:direct_injection_results} are
computed in a box of comoving side length $200\,\mathrm{cMpc}$, with density and ionization fields of $100^3$ cells, which suffices for volume-averaged quantities. The power spectra of section~\ref{sec:power_spectrum}, which require a larger volume to sample the largest scales, are computed in a box
of $600\,\mathrm{cMpc}$, with high-resolution halo grids of $800^3$ cells and density and ionization fields of $200^3$ cells. Table~\ref{tab:sims} lists the runs shown in the figures, with their parameters, heating channels and boxes.

\begin{table}
\centering
\caption{Simulations used in this work. All runs share the same initial
conditions and adopt the EOS21 astrophysical parameters~\citep{Munoz2022}
unless noted; CR models include no X-ray heating, X-ray models no CR heating,
and all include Coulomb losses. $\eta_{\rm cr}$ is in
$\mathrm{erg}\,M_\odot^{-1}$, $L_{\rm X}$ is $\log_{10}[L_{\rm X}/(\mathrm{erg\,s^{-1}}\,M_\odot^{-1}\,\mathrm{yr})]$ for the soft-band ($<2\,\mathrm{keV}$) luminosity,
$M_{\rm cr}^{\rm max}$ is in $M_\odot$, and the box side length is in cMpc.}
\label{tab:sims}
\footnotesize
\setlength{\tabcolsep}{4pt}
\begin{tabular}{lccccccc}
\hline
Model & $\eta_{\rm cr}$ & $\alpha$ & $L_{\rm X}$ &
$M_{\rm cr}^{\rm max}$ & Ly$\alpha$/CMB & Box & Figures \\
\hline
No heating          & --        & --  & --   & -- & off & 200 & \ref{fig:global_signal_eos21} \\
No CR or X-ray      & --        & --  & --   & -- & on  & 200 & \ref{fig:gas_temperature_comparison}, \ref{fig:gj_decomposition} \\
X-ray only          & --        & --  & 40.5 & -- & on  & 200, 600 & \ref{fig:gas_temperature_comparison}, \ref{fig:power_spectrum_vs_z} \\
                    & --        & --  & 40.5 & -- & off & 200 & \ref{fig:global_signal_eos21} \\
X-ray, matched$^a$  & --        & --  & 38.76--39.82 & -- & on & 600 & \ref{fig:ps_matched} \\
CR Pessimistic      & $10^{47}$ & 4.2 & --   & -- & on  & 200, 600 & \ref{fig:gas_temperature_comparison}, \ref{fig:power_spectrum_vs_z} \\
CR Fiducial         & $10^{48}$ & 4.2 & --   & -- & on  & 200, 600 & \ref{fig:gas_temperature_comparison}, \ref{fig:power_spectrum_vs_z}, \ref{fig:ps_matched} \\
                    & $10^{48}$ & 4.2 & --   & -- & off & 200 & \ref{fig:global_signal_eos21} \\
CR Optimistic       & $10^{49}$ & 4.2 & --   & -- & on  & 200, 600 & \ref{fig:sed}, \ref{fig:gas_temperature_comparison}, \ref{fig:gj_decomposition}, \ref{fig:power_spectrum_vs_z}, \ref{fig:ps_matched} \\
                    & $10^{49}$ & 4.2 & --   & -- & off & 200 & \ref{fig:global_signal_eos21} \\
GJ23 variants$^b$   & $10^{49}$ & 4.2, $E_{\rm K}^{-2}$ & -- & -- & on & 200 & \ref{fig:gj_decomposition} \\
MCG-only            & $10^{49}$ & 4.2 & --   & $10^{6}$--$10^{7}$ & on & 200 & \ref{fig:minihalo} (left) \\
                    & $(1.0$--$1.6)\times10^{51}$ & 4.2 & -- & $1.6\times10^{6}$ & on & 200 & \ref{fig:minihalo} (right) \\
\hline
\end{tabular}

\smallskip
\raggedright\footnotesize
$^a$\,$L_{\rm X}/\mathrm{SFR}$ retuned at each redshift to match the global
signal of the Optimistic or Fiducial CR model (figure~\ref{fig:ps_matched}).
$^b$\,Optimistic model with a constant stellar fraction
($\alpha_\star^{(\rm II)} = 0$), with the kinetic-energy power law
$\mathrm{d}N/\mathrm{d}E_{\rm K} \propto E_{\rm K}^{-2}$, or with both; the
no-CR model is also run with a constant stellar fraction.
\end{table}

% ============================================================
\section{Results}\label{sec:Results}
% ------------------------------------------------------------

We now use these simulations to quantify how much, and where, CRs heat the IGM, and whether their imprint on the 21-cm signal can be distinguished from that of X-rays. We start by identifying the CR energies responsible for the heating, as they determine how sensitive the heating is to the shape of the injection spectrum.

\subsection{Which protons heat the IGM}\label{sec:heating_rate_per_energy_bin}

To identify which CR protons heat the IGM, we decompose the heating rate of eq.~\eqref{eq:cr_spectral_integration} by the initial kinetic energy of the protons, defining the heating rate per unit logarithmic interval of initial energy,
\begin{equation}
\label{eq:heating_rate_per_bin}
\mathcal{H}(\mathbf{x}, E_{\rm K,i}, z') \equiv
\frac{\mathrm{d}\varepsilon_{\rm CR}}{\mathrm{d}\ln E_{\rm K,i}}
= \frac{E_{\rm K,i}}{n_{\rm b}(\mathbf{x}, z')}
\int_{z'}^{z''_{\max}} \mathrm{d}z''\,
\frac{\mathrm{d}^2 n_{\rm CR}}{\mathrm{d}E_{\rm K,i}\,\mathrm{d}z''}\,
\mathcal{L}\big[E_{\rm K}(E_{\rm K,i}; z', z'')\big] ,
\end{equation}
so that $\varepsilon_{\rm CR} = \int \mathcal{H}\,\mathrm{d}\ln E_{\rm K,i}$. We compute $\mathcal{H}$ in every cell, average it over a subsample of cells, and convert it into a temperature change per unit redshift and per unit $\ln E_{\rm K,i}$ with the first term of eq.~\eqref{eq:tk_temperature_update}, $|\mathrm{d}^2 T_{\rm K}/\mathrm{d}z'\,\mathrm{d}\ln E_{\rm K,i}| = 2\,|\mathrm{d}t/\mathrm{d}z'|\,\langle\mathcal{H}\rangle/[3k_{\rm B}(1+x_{\rm e})]$.

\begin{figure}
    \centering
        \includegraphics[width=0.78\columnwidth]{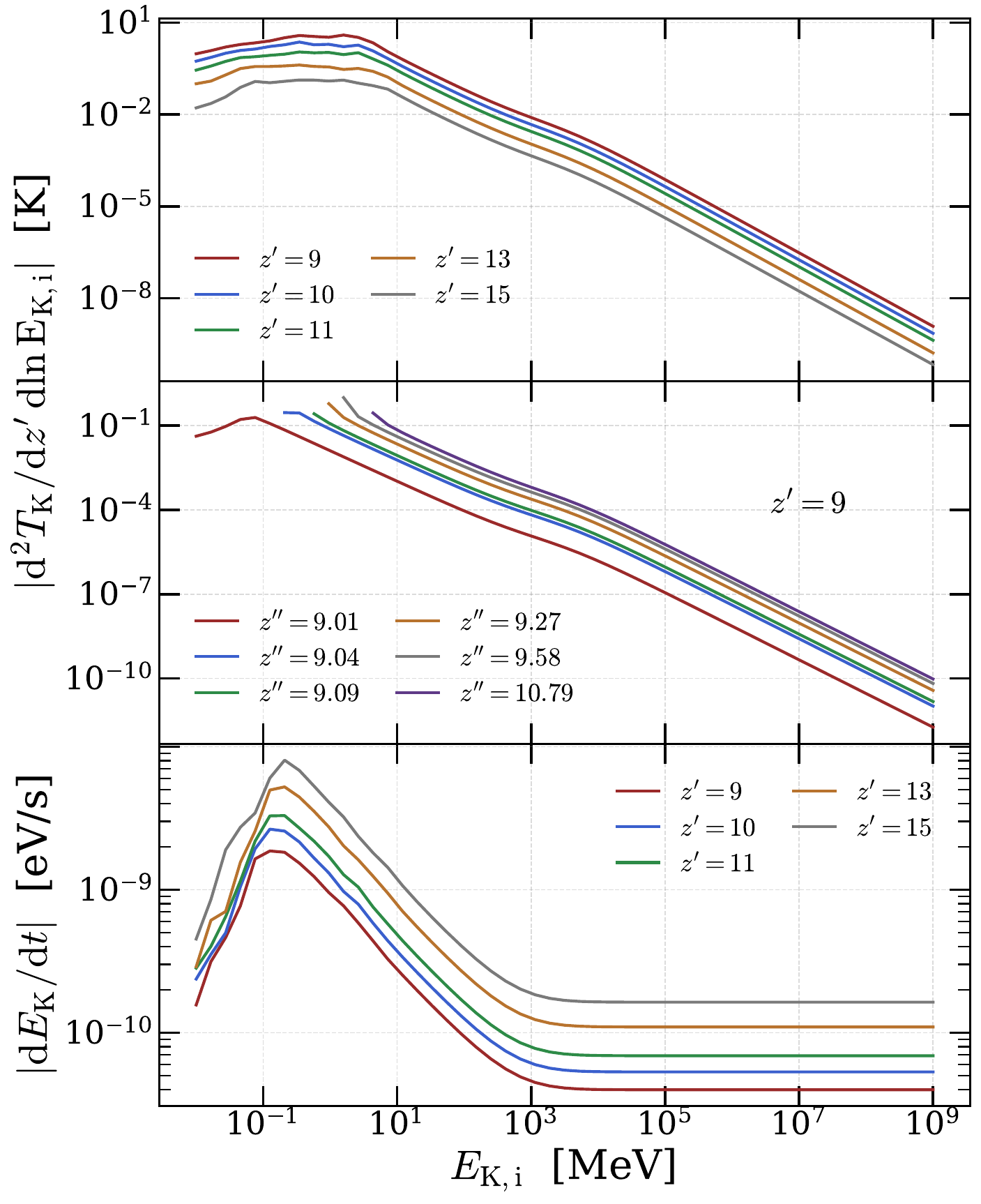}
    \caption{Energy dependence of CR heating in the optimistic model
    ($\eta_{\rm cr} = 10^{49}\,\mathrm{erg}\,M_\odot^{-1}$, $\alpha = 4.2$), computed on 50 logarithmic bins in $E_{\rm K,i}$ between $0.01$ and $10^{9}\,\mathrm{MeV}$. \textit{Top:} heating rate per unit logarithmic
    interval of initial kinetic energy,
    $|\mathrm{d}^2T_{\rm K}/\mathrm{d}z'\,\mathrm{d}\ln E_{\rm K,i}|$, at
    $9 \leq z' \leq 15$, summed over source shells. \textit{Middle:} the  $z' = 9$ rate separated by source shell, labeled by emission redshift $z''$; protons from more distant shells reach the cell only above a threshold energy, below which they thermalize in transit. \textit{Bottom:} collisional (excitation, ionization and Coulomb) energy-loss rate $|\mathrm{d}E_{\rm K}/\mathrm{d}t|$, excluding adiabatic losses, evaluated at the arrival energy of a proton of initial energy $E_{\rm K,i}$ and averaged over source shells, at the same redshifts as the top panel. It is dominated by excitation and ionization, Coulomb losses off free electrons contributing well under $1\%$ in the largely neutral gas, and its maximum lies at $E_{\rm K,i} \approx 0.1$--$0.2\,\mathrm{MeV}$ rather than at the $47\,\mathrm{keV}$ Bethe--Bloch peak of eq.~\eqref{eq:bethe_bloch}, which refers to the arrival energy. Heating rates are averaged over a subsample of simulation cells.}
    \label{fig:sed}
\end{figure}

The top panel of figure~\ref{fig:sed} shows the resulting heating rate per
unit $\ln E_{\rm K,i}$ at $9 \leq z' \leq 15$. At all redshifts it forms a
broad plateau, within a factor of two of its maximum between $\approx 0.1$ and $\approx 4\,\mathrm{MeV}$ at $z' = 9$, and falls as a power law toward higher energies. This shape is the product of the injection spectrum, which falls with energy, and the collisional energy-loss rate (bottom panel), which is dominated by excitation and ionization, peaks at
$E_{\rm K,i} \approx 0.1$--$0.2\,\mathrm{MeV}$, and declines as $1/\beta \propto E_{\rm K}^{-1/2}$ above the Bragg peak, flattening at relativistic energies. At higher redshift, the loss rate is larger, as the IGM is denser ($n_{\rm b} \propto (1+z)^3$), but the number density of CR protons is smaller, as the SFRD has had less time to build up. The latter effect wins: the heating rate decreases monotonically from $z' = 9$ to $z' = 15$ at every energy, while keeping a similar shape. Integrated over $\ln E_{\rm K,i}$, the $z' = 9$ rate agrees to within $10\%$ with the CR heating rate implied by the temperature evolution of the optimistic model in figure~\ref{fig:gas_temperature_comparison}.

In the middle panel of figure~\ref{fig:sed}, we decompose the $z' = 9$ heating rate into contributions from individual source shells, labeled by their emission redshift $z''$, for six shells logarithmically spaced in comoving radius between $z'' \approx 9.01$ and $z'' \approx 10.8$. Only the nearest shells contribute at the lowest initial energies, since CRs from these shells have had little time to lose energy in transit. More distant shells contribute only above an initial energy that rises from $\approx 0.2\,\mathrm{MeV}$ for $z'' = 9.04$ to $\approx 4\,\mathrm{MeV}$ for $z'' \approx 10.8$: following eq.~\eqref{eq:energy_evolution}, CRs injected at lower energies at higher $z''$ lose all of their kinetic energy before reaching $z' = 9$. Above a few MeV, where protons from every shell arrive, the more distant shells dominate, as they enclose a larger volume and hence more sources. This cutoff is the thermalization horizon retained by our transport approximation (section~\ref{sec:implementation}): although each shell is assigned its light-travel redshift, protons that thermalize in transit contribute nothing to the heating.
%The threshold energy below which a CR thermalizes in transit
%increases with $z''$, which is why the cliff shifts to higher
%$E_{\rm{K},i}$ for more distant shells. At energies where all shells
%contribute, however, the higher $z''$ curves sit above the
%lower $z''$ curves: distant shells deliver more energy per energy
%bin than nearby ones because they sample a larger comoving volume of
%source galaxies. The dominance of nearby shells at the lowest energies
%and distant shells at the highest energies together produce the total
%heating rate shown in the top panel.

\subsection{IGM thermal history}\label{sec:tk_evolution}

\begin{figure}
    \centering
    \includegraphics[width=\columnwidth]{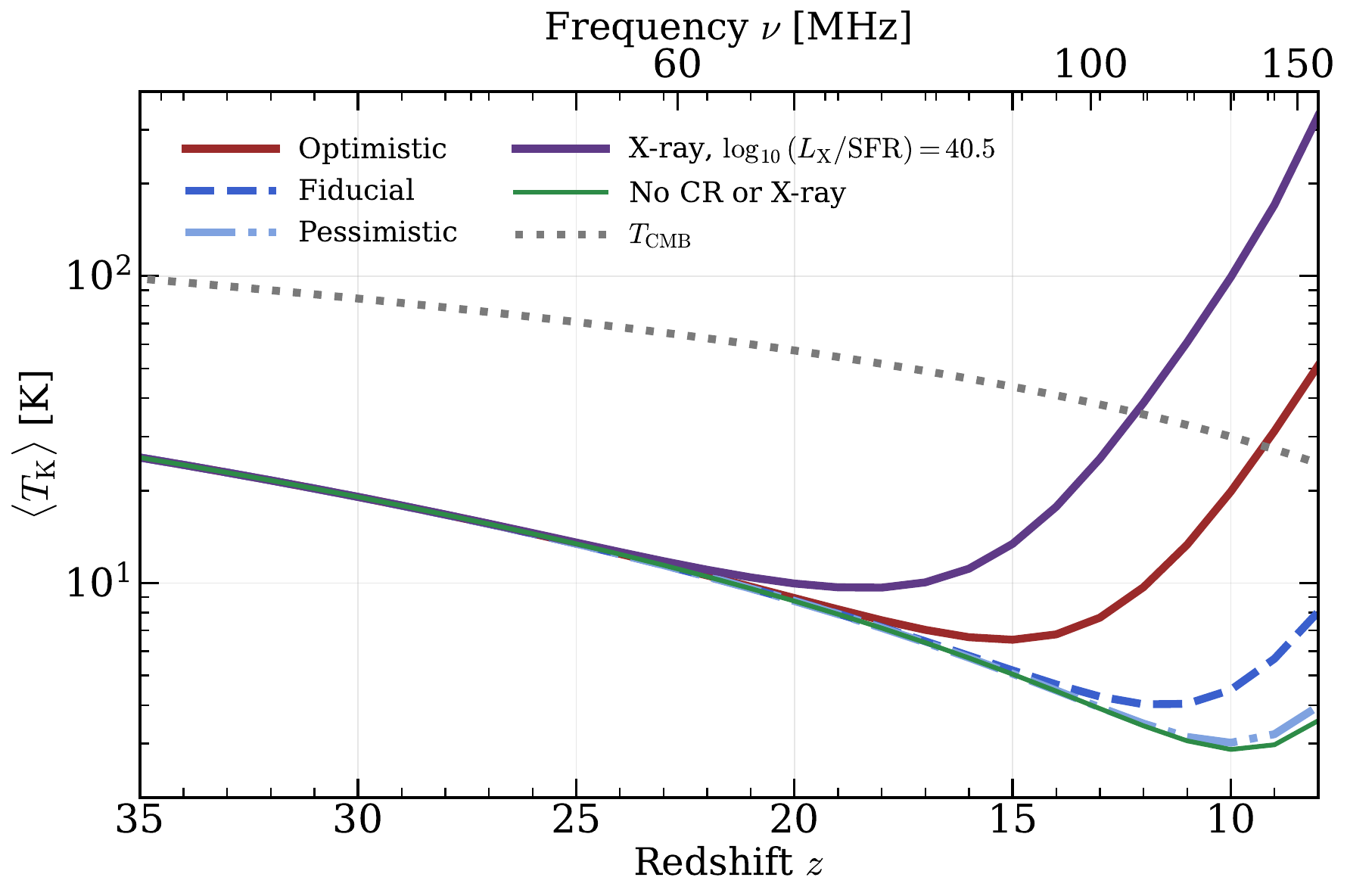}
    \caption{Volume-averaged IGM kinetic temperature, $\langle T_{\rm K}\rangle$, for CR heating with $\alpha = 4.2$ and three efficiencies: optimistic ($\eta_{\rm cr} = 10^{49}\,\mathrm{erg}\,M_\odot^{-1}$; solid red curve), fiducial ($10^{48}\,\mathrm{erg}\,M_\odot^{-1}$; dashed blue curve) and pessimistic ($10^{47}\,\mathrm{erg}\,M_\odot^{-1}$; dash-dotted light blue curve). For reference, the purple curve shows the X-ray-only reference model, with $\log_{10}[L_{\rm X}/(\mathrm{erg\,s^{-1}}\,M_\odot^{-1}\,\mathrm{yr})] = 40.5$, the thin green curve the model without CR or X-ray heating, and the dotted gray curve the CMB temperature. All models include Ly$\alpha$ and CMB heating, which account for the departure of the green curve from adiabatic cooling below $z \approx 12$. The upper axis gives the observed frequency, $\nu = \nu_{21}/(1+z)$.}
    \label{fig:gas_temperature_comparison}
\end{figure}

Having identified which CR energies heat the IGM, we now follow the gas temperature (figure~\ref{fig:gas_temperature_comparison}) in the pessimistic, fiducial and optimistic CR models, which bracket the range explored by GJ23, and in the X-ray-only reference model of section~\ref{sec:sims}, all including Ly$\alpha$ and CMB heating. Compton heating by the residual free electrons is retained in all runs but is negligible below $z \sim 150$, so that gas heated by none of these channels would cool adiabatically, as $T_{\rm K} \propto (1+z)^2$ \citep{Furlanetto2006review, PritchardLoeb2012}.

We find that even our optimistic CR model heats the IGM considerably less
than the X-ray-only reference model. With
$\eta_{\rm cr} = 10^{49}\,\mathrm{erg}\,M_\odot^{-1}$, $T_{\rm K}$ reaches
its minimum of $\approx 6.5\,\mathrm{K}$ near $z \approx 15$ and exceeds
$T_{\rm CMB}$ only at $z \lesssim 9$, whereas X-ray heating at the EOS21
fiducial luminosity reaches its minimum of $\approx 9.7\,\mathrm{K}$ already
near $z \approx 18$ and exceeds $T_{\rm CMB}$ by $z \approx 12$. The two
histories are therefore offset by $\Delta z \approx 3$ throughout the
heating epoch, comparable to the offsets of $\Delta z \approx 2$--$4$
between the absorption troughs of the global signal
(section~\ref{sec:global_signal_eos21}), and by $z \approx 8$ the X-ray
model is about six and a half times as hot as the optimistic CR model
($\approx 330$ against $\approx 50\,\mathrm{K}$). Even at the highest
efficiency we consider, CR heating thus results in a later and shallower
thermal history than X-ray heating at its fiducial luminosity.

As $\eta_{\rm cr}$ decreases, heating starts later and is weaker. The
fiducial model reaches a minimum of $T_{\rm K} \approx 4\,\mathrm{K}$ near
$z \approx 11$--$12$ and rises only to $\approx 8\,\mathrm{K}$ by $z = 8$,
while the pessimistic model stays within $1\%$ of the model without CR or
X-ray heating above $z \approx 13$ and exceeds it by only $0.4\,\mathrm{K}$
at $z = 8$. Across two orders of magnitude in $\eta_{\rm cr}$, CR heating
with the momentum power law thus ranges from a transition delayed by
$\Delta z \approx 3$ relative to X-ray heating (optimistic), through a later
and weaker one (fiducial), to a negligible contribution (pessimistic). None of these comparisons is sensitive to Coulomb losses off free electrons, which are included in all runs but raise $\langle T_{\rm K}\rangle$ by less than $0.5\%$ at $z \gtrsim 13$ and by $\sim 1.8\%$ by $z \sim 6$.

That CR heating remains weaker than X-ray heating even at $\eta_{\rm cr} = 10^{49}\,\mathrm{erg}\,M_\odot^{-1}$ contrasts with GJ23, who found CR heating at $\eta_{\rm cr} = 10^{48}\,\mathrm{erg}\,M_\odot^{-1}$ to be roughly as efficient as, and slightly more efficient than, X-ray heating with $f_{\rm X} = 1$ at $z < 20$ (their section~4.3). Two differences account for this. First, our momentum power law moves CR energy out of the sub-$40\,\mathrm{MeV}$ range that dominates IGM heating (section~\ref{sec:attribution}), reducing the heat deposited per unit injected CR energy. Second, our X-ray benchmark is stronger: the EOS21 fiducial soft-band ($<2\,\mathrm{keV}$) $L_{\rm X}/\mathrm{SFR} = 10^{40.5}\,\mathrm{erg\,s^{-1}}\,M_\odot^{-1}\,\mathrm{yr}$ corresponds to $f_{\rm X} > 1$ in their notation, in which $f_{\rm X} = 1$ denotes a bolometric $L_{\rm X}/\mathrm{SFR} = 3\times10^{40}\,\mathrm{erg\,s^{-1}}\,M_\odot^{-1}\,\mathrm{yr}$ with the X-ray binary spectrum of \citet{Fragos2013} (see~\citep[eq.~1]{Fialkov2014}).

\subsection{Why cosmic-ray heating is weaker than previously estimated}\label{sec:attribution}

\begin{figure}
    \centering
    \includegraphics[width=\columnwidth]{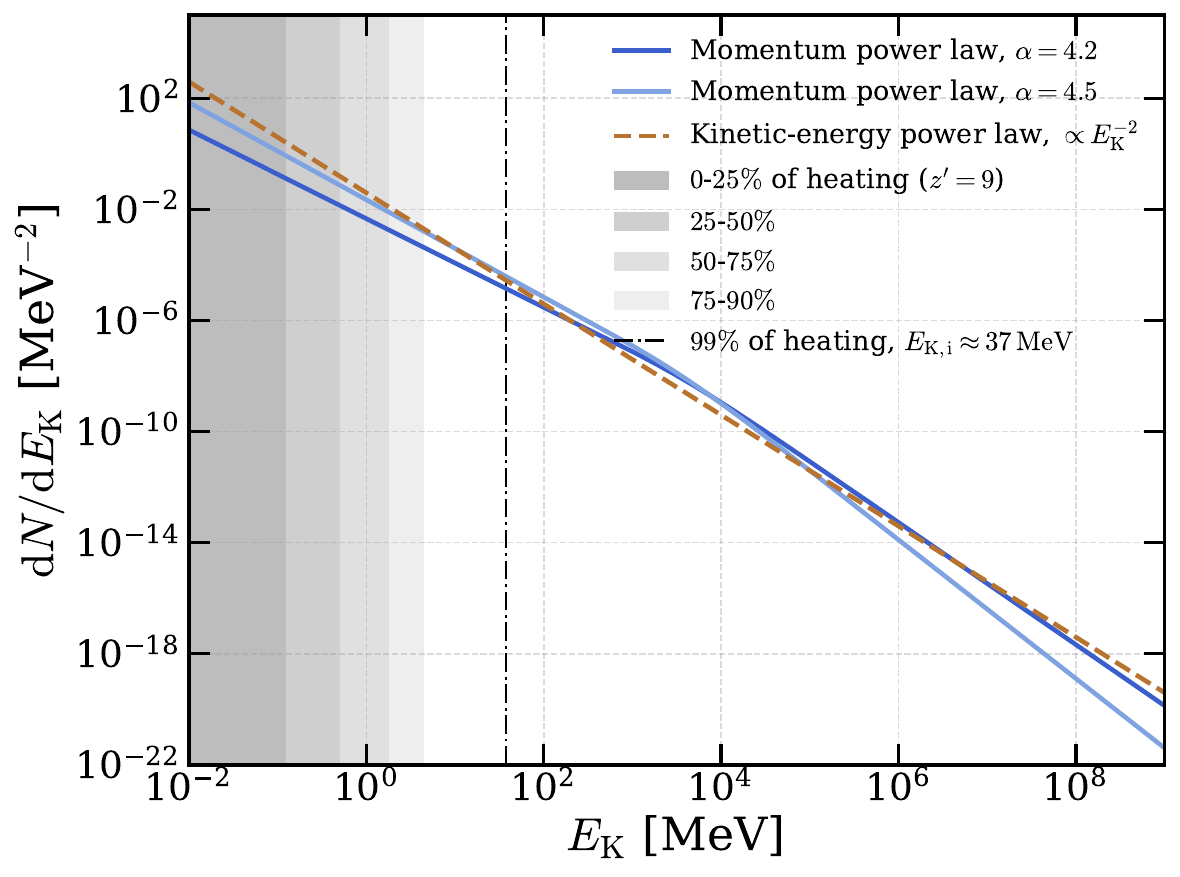}
    \caption{Injection spectra normalized to unit total kinetic energy between
$E_{\rm K,min} = 0.01$ and $E_{\rm K,max} = 10^{9}\,\mathrm{MeV}$
(eq.~\eqref{eq:shape_norm}): momentum power laws (this work) with
$\alpha = 4.2$ (blue) and $\alpha = 4.5$ (light blue), and the
kinetic-energy power law of \citet{GesseyJones2023},
$\mathrm{d}N/\mathrm{d}E_{\rm K} \propto E_{\rm K}^{-2}$ (dashed brown). Gray bands, from darkest to lightest, are bounded by the energies below which $25$, $50$, $75$ and $90\%$ of the CR heating at $z' = 9$ originates, and the vertical dash-dotted line marks the $99\%$ threshold, $E_{\rm K,i} \approx 37\,\mathrm{MeV}$, all computed from the heating rate
per unit $\ln E_{\rm K,i}$ (figure~\ref{fig:sed}). Within the
heating-dominated range, the kinetic-energy power law lies above the
momentum power laws, which instead place more energy at
$10^{2} \lesssim E_{\rm K} \lesssim 10^{6}\,\mathrm{MeV}$, where the IGM is
largely transparent to CRs.}
    \label{fig:spectrum_comparison}
\end{figure}

At matched $\eta_{\rm cr}$, our CR heating is more than an order of magnitude weaker than that of GJ23, primarily because of the injection spectrum. The momentum and kinetic-energy power laws are within a factor of three of each other at intermediate energies ($10^2 \lesssim E_{\rm K} \lesssim 10^4\,\mathrm{MeV}$) but diverge at low energies (figure~\ref{fig:spectrum_comparison}): in the non-relativistic limit, our spectrum reduces to $\mathrm{d}N/\mathrm{d}E_{\rm K} \propto E_{\rm K}^{(1-\alpha)/2}$, which is shallower than the $\mathrm{d}N/\mathrm{d}E_{\rm K} \propto E_{\rm K}^{-2}$ of GJ23. The kinetic-energy power law therefore places substantially more energy at $E_{\rm K} \lesssim 10\,\mathrm{MeV}$, the regime that dominates IGM heating.

\begin{figure}
\centering
\includegraphics[width=\columnwidth]{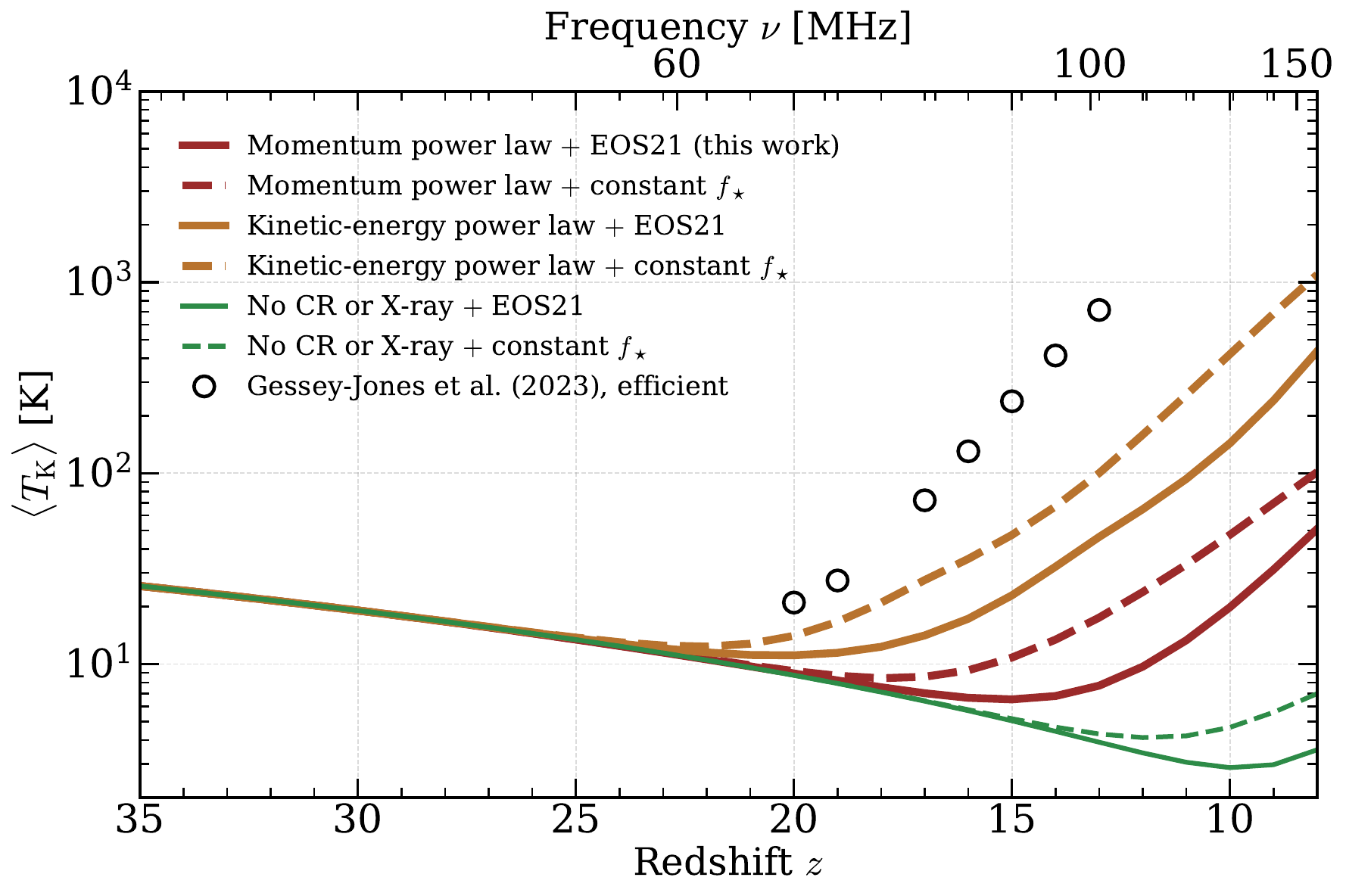}
\caption{Decomposition of the reduced CR heating relative to
\citet{GesseyJones2023}, at the matched efficiency
$\eta_{\rm cr} = 10^{49}\,\mathrm{erg}\,M_\odot^{-1}$. Color distinguishes
the injection spectrum, our momentum power law with $\alpha = 4.2$ (dark
red) or the kinetic-energy power law of \citet{GesseyJones2023},
$\mathrm{d}N/\mathrm{d}E_{\rm K} \propto E_{\rm K}^{-2}$ (brown); line style
distinguishes the galaxy model, EOS21 (solid) or a constant stellar
fraction, $\alpha_\star^{(\rm II)} = 0$, matched to the EOS21 relation at
the pivot mass $10^{10}\,M_\odot$ (dashed). The solid dark red curve is our
optimistic model. Green curves show the corresponding models without CR or
X-ray heating; as in \citet{GesseyJones2023}, all models include Ly$\alpha$
and CMB heating. Open circles mark the efficient model of \citet{GesseyJones2023}, digitized from their figure~7 ($\eta_{\rm cr} = 10^{49}\,\mathrm{erg}\,
M_\odot^{-1}$, $\mathrm{d}N/\mathrm{d}E_{\rm K} \propto E_{\rm K}^{-2}$). The injection
spectrum, rather than the galaxy model, is the dominant cause of the reduced
heating.}
\label{fig:gj_decomposition}
\end{figure}

The gray bands in figure~\ref{fig:spectrum_comparison} show the cumulative contribution to the heating at $z' = 9$ from CRs below a given initial energy (from the top panel of figure~\ref{fig:sed}): half of the heating comes from $E_{\rm K,i} \lesssim 0.5\,\mathrm{MeV}$, and $99\%$ from $E_{\rm K,i} \lesssim 37\,\mathrm{MeV}$, close to the $30\,\mathrm{MeV}$ boundary GJ23 use when comparing injection spectra. Our momentum power law places only $4.3\%$ of the injected CR energy below
$30\,\mathrm{MeV}$ and $0.7\%$ below $0.5\,\mathrm{MeV}$, compared with $32\%$ and $15\%$ for the kinetic-energy power law, i.e., $7$ and $22$ times less; the factor of $\approx 12$ in heating found below lies between the
two, since the heating draws on both energy ranges.

The heating is also sensitive to the spectral index: steeper spectra (larger $\alpha$) place more energy at low energies (compare the $\alpha = 4.2$ and $4.5$ curves in figure~\ref{fig:spectrum_comparison}); an $\alpha = 4.5$ spectrum places $18\%$ of the injected energy below $30\,\mathrm{MeV}$, against $4.3\%$ for $\alpha = 4.2$, and therefore heats the IGM more. Across the range inferred for Galactic remnants, $\alpha \approx 4.2$--$4.4$, the momentum power law places between $7$ and $2.6$ times less energy below $30\,\mathrm{MeV}$ than the kinetic-energy power law. We adopt $\alpha = 4.2$ as our fiducial value, motivated by the slightly steeper-than-test-particle spectra inferred for Galactic supernova remnants, e.g.~\citep{Vink2012, Caprioli2012}. A quantitative comparison with GJ23 therefore depends on both the spectral form and the value of $\alpha$.

To separate the contributions of the injection spectrum and of the galaxy
model, we repeat our optimistic model replacing each ingredient in turn
(figure~\ref{fig:gj_decomposition}). Measuring the heating as the excess of
$\langle T_{\rm K} \rangle$ over the corresponding model without CR
heating, which, with Ly$\alpha$ and CMB heating enabled, lies above the
adiabatic curve, we find at $z = 15$ that replacing the EOS21 SHMR with a
constant stellar fraction raises the heating by a factor of $3.8$, replacing
our momentum power law with the kinetic-energy power law by a factor of
$12$, and replacing both by a factor of $28$. The two effects partly
overlap, their product ($\approx 46$) exceeding the combined factor, but
the injection spectrum is the larger of the two by a factor of $\approx 3$.
The spectral factor lies within the shift in spectral weight quantified
above (seven times more energy below $30\,\mathrm{MeV}$, and $22$ times more below $0.5\,\mathrm{MeV}$, in the kinetic-energy power law), so it follows
from the energy made available at sub-relativistic energies rather than
from any difference in transport.

Replacing both ingredients brings our thermal history to within a factor of $2.6$ in $T_{\rm K}$ of that of GJ23 at $z = 17$, with a residual that grows to $\approx 7$ by $z = 13$, the lowest redshift shown in their figure~7. The remainder likely owes to the normalization of the constant stellar fraction, which we match to the EOS21 relation at $10^{10}\,M_\odot$ rather than to the value adopted by GJ23, and to the remaining differences between the two galaxy models.

% ------------------------------------------------------------

\subsection{Cosmic-ray versus X-ray heating in the global signal}\label{sec:global_signal_eos21}

\begin{figure}
    \centering
    \includegraphics[width=0.75\columnwidth]{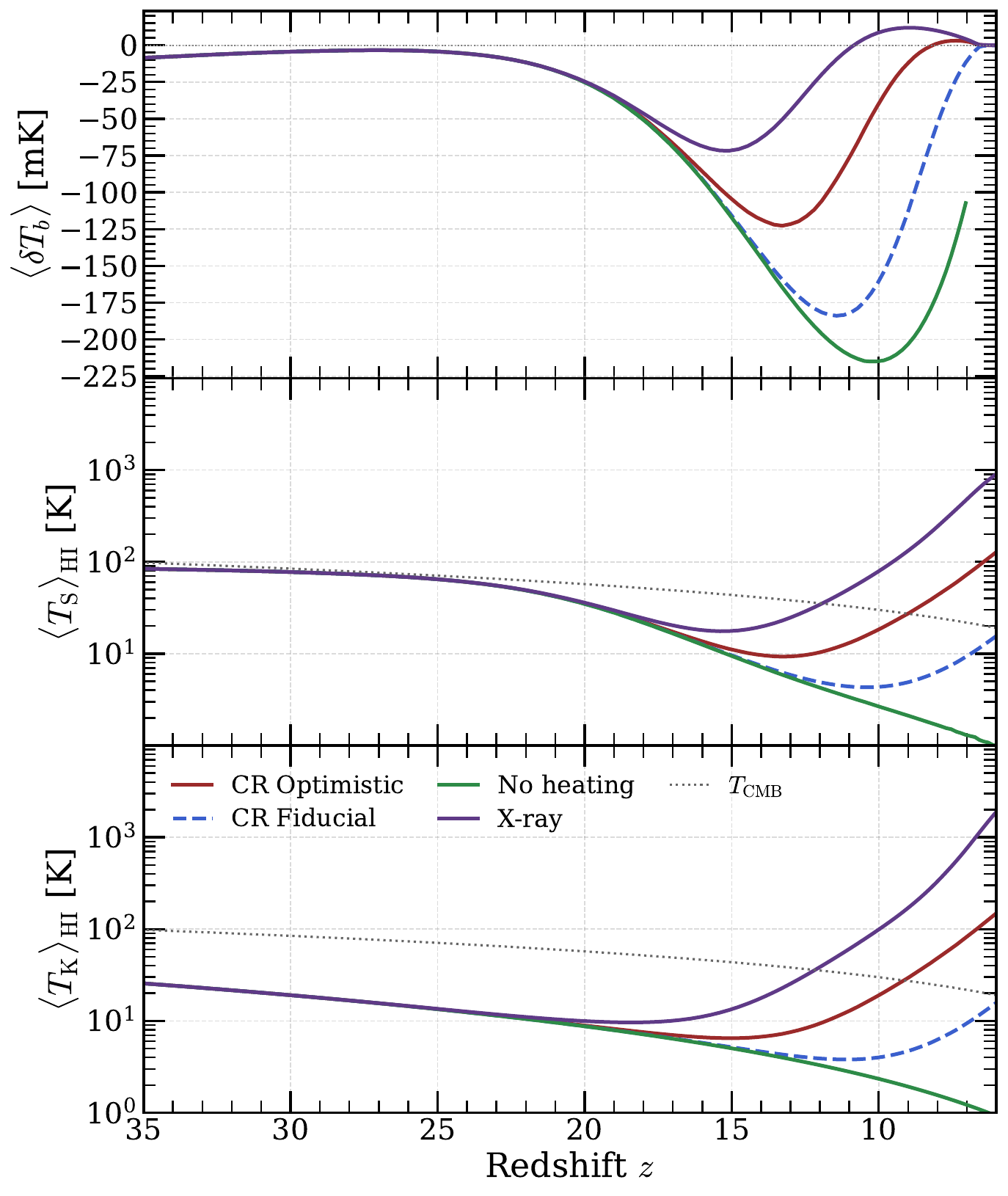}
    \caption{Global 21-cm signal (top panel), and spin (middle) and kinetic (bottom) temperatures averaged over the neutral IGM, for four models without Ly$\alpha$ and CMB heating: no heating (green), X-ray only with $L_{\rm X}/\mathrm{SFR} = 10^{40.5}\,\mathrm{erg\,s^{-1}}\,M_\odot^{-1}\,\mathrm{yr}$ (purple), and CR heating with $\alpha = 4.2$ at the fiducial ($\eta_{\rm cr} = 10^{48}\,\mathrm{erg}\,M_\odot^{-1}$; dashed blue) and optimistic ($10^{49}\,\mathrm{erg}\,M_\odot^{-1}$; solid red) efficiencies. Temperatures are averaged over the neutral phase because ionized regions in \texttt{21cmFAST} carry a separate temperature field. Without Ly$\alpha$ and CMB heating, the temperatures at $z \lesssim 12$ are lower than in figure~\ref{fig:gas_temperature_comparison}, where both are included. Dotted gray curves show $T_{\rm CMB}$. The no-heating signal is truncated at $z = 7$, where reionization drives it off the plotted range. At these efficiencies, the X-ray trough occurs $\Delta z \approx 2$ (optimistic) and $\approx 4$ (fiducial) earlier than the CR troughs.}

    \label{fig:global_signal_eos21}
\end{figure}

We now compare the global signals of CR and X-ray heating for CR injection from both ACGs and MCGs at a common $\eta_{\rm cr}$, using four simulations on a $200\,\mathrm{cMpc}$ box without Ly$\alpha$ and CMB heating: the no-heating model, the X-ray-only model at the EOS21 fiducial $L_{\rm X}/\mathrm{SFR} = 10^{40.5}\,\mathrm{erg\,s^{-1}}\,M_\odot^{-1}\,\mathrm{yr}$ \citep{Munoz2022}, and the fiducial and optimistic CR models. In figure~\ref{fig:global_signal_eos21} we show the global signal $\langle \delta T_b \rangle (z)$ (top panel), and the spin (middle) and kinetic (bottom) temperatures of the neutral IGM. The reionization histories of the four runs (not shown) are indistinguishable, as the heating channel does not affect the ionizing photon budget and the partial ionization by CRs is negligible at these efficiencies.

As in section~\ref{sec:tk_evolution}, X-ray heating at the fiducial $L_{\rm X}/\mathrm{SFR}$ lifts the IGM above $T_{\rm CMB}$ by $z \approx 12$, whereas the optimistic CR model does so only at $z \lesssim 9$ and the fiducial one not at all (bottom panel of figure~\ref{fig:global_signal_eos21}). The X-ray global signal therefore reaches an absorption minimum of $-72\,\mathrm{mK}$ at $z \approx 15$, turns into emission at $z \approx 11$, and peaks at $\approx +12\,\mathrm{mK}$ near $z \approx 9$, while that of the optimistic CR model turns into weak emission only below $z \approx 8$, peaking at $\approx 3\,\mathrm{mK}$ near $z \approx 7.4$. The optimistic model
reaches an absorption minimum of $\approx -123\,\mathrm{mK}$ at $z \approx 13.3$
and the fiducial model a deeper minimum of $\approx -184\,\mathrm{mK}$ at
$z \approx 11.4$, compared with $-215\,\mathrm{mK}$ at $z \approx 10.3$ in the
no-heating model. Optimistic and fiducial CR heating thus make the trough
shallower by $\approx 92$ and $\approx 31\,\mathrm{mK}$, respectively.

The troughs also differ in redshift. At the fiducial $L_{\rm X}/\mathrm{SFR}$ of \citet{Munoz2022}, X-rays deliver enough energy by $z \approx 15$ for the signal to start climbing out of absorption, whereas the CR troughs occur at $z \approx 13.3$ (optimistic) and $z \approx 11.4$ (fiducial), i.e., $\Delta z \approx 2$--$4$ later, because the momentum power law with $\alpha = 4.2$ deposits less heat per unit star formation (section~\ref{sec:tk_evolution}). At these literature efficiencies, the trough position alone would thus separate the two channels. Coulomb losses do not change this picture: in the optimistic CR model they make the trough shallower by only $\sim 0.7\,\mathrm{mK}$ and shift it negligibly in redshift, their largest imprint being a $\sim 1.3\,\mathrm{mK}$ rise on the recovering side of the trough, near $z \approx 10$.

Both efficiencies are, however, uncertain by orders of magnitude. Once $\eta_{\rm cr}$ and $L_{\rm X}/\mathrm{SFR}$ are treated as free parameters, the global 21-cm signal alone cannot distinguish X-ray from CR heating: a sufficiently large $\eta_{\rm cr}$ can reproduce the absorption depth of the X-ray model, and the trough redshift can be matched by retuning either channel. Indeed, the X-ray luminosities that reproduce the global signal of the optimistic (fiducial) CR model at $z = 16$, $14$, $12$ and $10$ differ by at most $0.06$ ($0.09$)\,dex (section~\ref{sec:power_spectrum}; these matched runs include Ly$\alpha$ and CMB heating, unlike those of figure~\ref{fig:global_signal_eos21}), so that a single X-ray model would closely mimic each CR global signal. The required efficiencies are not freely available: matching the X-ray model at the EOS21 fiducial would require $\eta_{\rm cr} \approx 5\times10^{49}\,\mathrm{erg}\,M_\odot^{-1}$ (extrapolating the matched pairs linearly), above the ceiling set by Pop~II supernova energetics (section~\ref{sec:escape}), so that the degeneracy is bounded on one side by the energy budget rather than by the observable itself. Within the physically admissible range, breaking it requires either an independent prior on the energy injected per unit star formation, or the scale dependence of the 21-cm power spectrum, as the different ranges of the two heating carriers leave distinct spatial signatures \citep{GesseyJones2023}. We turn to the latter in section~\ref{sec:power_spectrum}.

\subsection{Power spectrum at matched global signal}\label{sec:power_spectrum}

The global signal collapses the heating field to a single number per redshift, discarding how the heat is distributed in space. Yet it is this distribution that should separate the two channels, as CRs deposit their energy closer to their sources than X-rays (sections~\ref{sec:losses} and~\ref{sec:architecture}). We therefore compare the 21-cm power spectra, $\Delta^2(k, z)$ (eq.~\eqref{eq:power_spectrum}), of the X-ray-only model and of the pessimistic, fiducial and optimistic CR models, computed from the brightness temperature box at each redshift in logarithmic bins of $k$. Unlike those of section~\ref{sec:global_signal_eos21}, these runs include Ly$\alpha$ and CMB heating, as appropriate for comparison with observations.

\begin{figure}
    \centering
    \includegraphics[width=0.95\columnwidth]{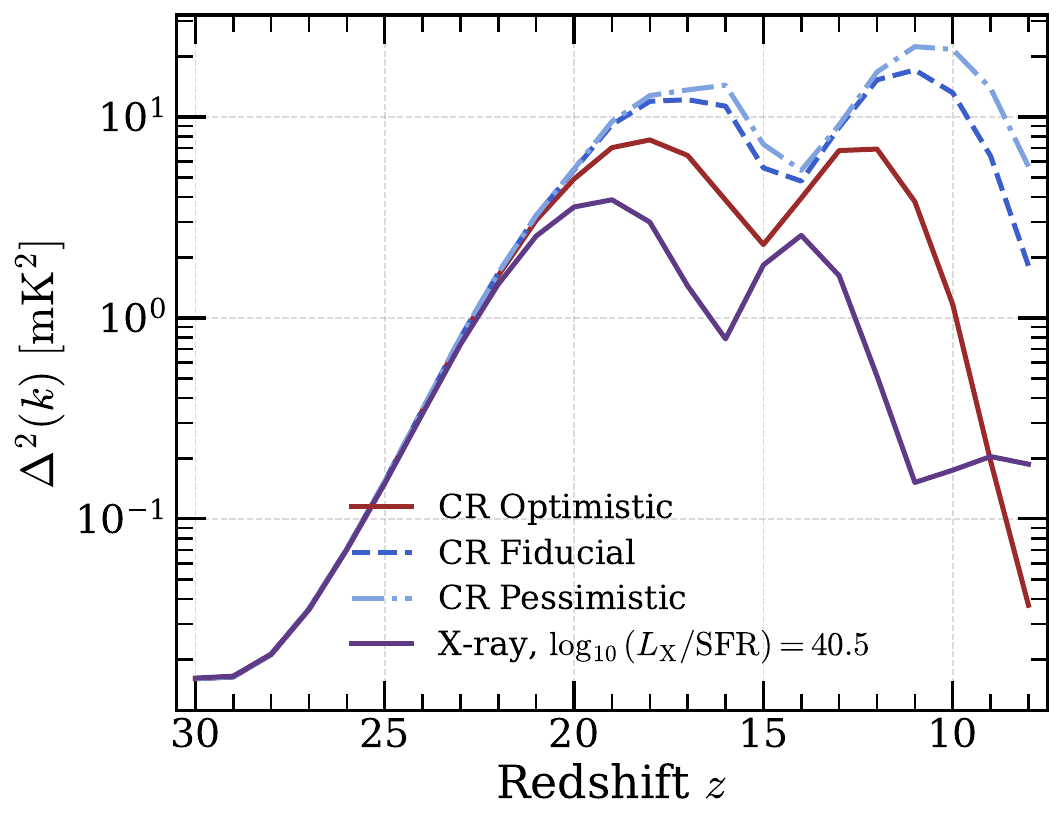}
    \caption{Redshift evolution of the 21-cm power spectrum at
$k = 0.16\,\mathrm{cMpc}^{-1}$ for the optimistic
($\eta_{\rm cr} = 10^{49}\,\mathrm{erg}\,M_\odot^{-1}$; solid red), fiducial
($10^{48}\,\mathrm{erg}\,M_\odot^{-1}$; dashed blue) and pessimistic
($10^{47}\,\mathrm{erg}\,M_\odot^{-1}$; dash-dotted light blue) CR models,
and for the X-ray-only model with $L_{\rm X}/\mathrm{SFR} =
10^{40.5}\,\mathrm{erg\,s^{-1}}\,M_\odot^{-1}\,\mathrm{yr}$ (purple), all
including Ly$\alpha$ and CMB heating. The CR heating peaks occur later than the X-ray one, with an amplitude that increases as $\eta_{\rm cr}$ decreases. This wavenumber lies near the VAO feature (section~\ref{sec:power_spectrum}), which affects all models alike.}
    \label{fig:power_spectrum_vs_z}
\end{figure}

In figure~\ref{fig:power_spectrum_vs_z} we show the evolution of the power at $k = 0.16\,\mathrm{cMpc}^{-1}$. All four models rise together from $z \sim 30$ to $z \approx 22$, as the power is sourced by fluctuations in the Ly$\alpha$ coupling, which are identical in runs that share the same star formation history; they separate once temperature fluctuations begin to contribute, in order of their heating onset. The X-ray power turns over first, at $z \approx 19$, passes through a minimum near $z \approx 16$, marking the transition from coupling-driven to temperature-driven fluctuations, and peaks at $z \approx 14$ before declining as the IGM saturates. In the CR models, heated later (section~\ref{sec:attribution}), the coupling-driven rise continues until $z \approx 16$--$18$ and the heating peak occurs at $z \approx 11$--$13$, the power-spectrum counterpart of the trough offset in the global signal.

Within the CR models, the peak amplitude increases as $\eta_{\rm cr}$ decreases, so that the pessimistic model produces the largest fluctuations. This inversion mostly reflects the mean contrast: the colder the IGM, the larger $|1 - T_{\rm CMB}/T_{\rm S}|$, which amplifies all fluctuations of $\delta T_b$, from density and ionization as well as temperature. At $z = 10$, the mean signal is $-152\,\mathrm{mK}$ in the pessimistic model and $-27\,\mathrm{mK}$ in the optimistic one (see below). At this wavenumber the CR power exceeds that of the X-ray-only model by up to a factor of $\approx 150$ (pessimistic model, $z \approx 11$). Coulomb losses change the heating of each cell by less than $1\%$ at the redshifts shown and leave the power spectra unchanged.

Comparing the two channels at their default efficiencies, however, conflates two distinct differences: in the amount of heat delivered, and in how that heat is distributed in space. At $z = 10$, for example, the X-ray model has already driven $\langle \delta T_b \rangle$ into emission, at $+8.8\,\mathrm{mK}$, while the CR models span $-27$ to $-152\,\mathrm{mK}$, so that most of the difference in amplitude reflects the different stages of thermal evolution the two channels have reached, rather than a property of the heating channel itself. To isolate the spatial signature, we retune $L_{\rm X}/\mathrm{SFR}$ at each redshift until the global signal of the X-ray model matches that of each CR model, and compare the power spectra of the resulting pairs.

\begin{figure*}
    \centering
    \includegraphics[width=\textwidth]{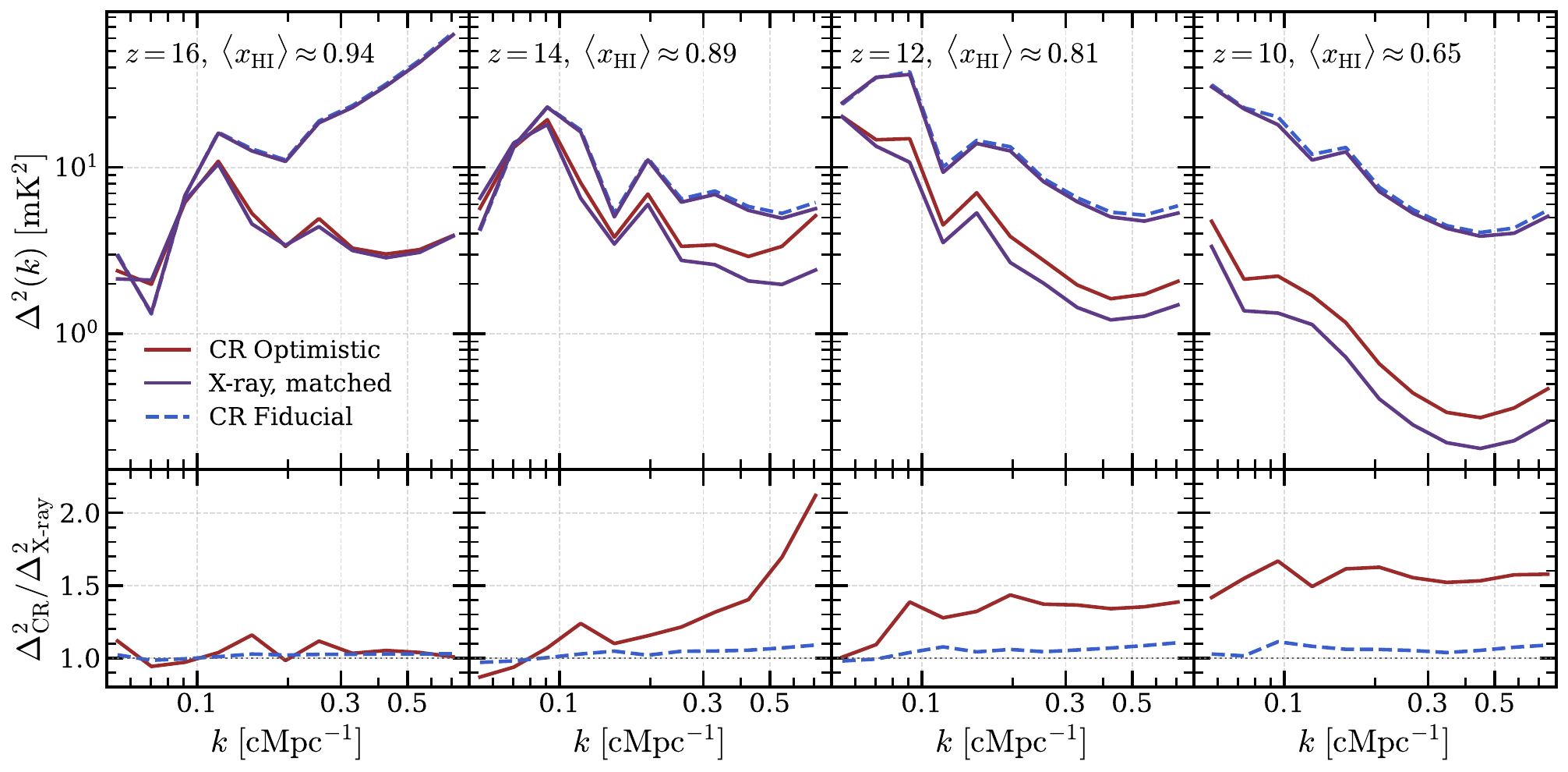}
    \caption{21-cm power spectra of CR and X-ray models matched in the global signal, at four redshifts during the heating epoch (\textit{left to right}). \textit{Top row:} $\Delta^2(k)$ for the optimistic ($\eta_{\rm cr} = 10^{49}\,\mathrm{erg}\,M_\odot^{-1}$; solid red) and fiducial ($10^{48}\,\mathrm{erg}\,M_\odot^{-1}$; dashed blue) CR models, each with the X-ray model that reproduces its $\langle \delta T_b \rangle$ (solid purple). \textit{Bottom row:} ratio $\Delta^2_{\rm CR}/\Delta^2_{\rm X\text{-}ray}$ for the optimistic (solid red) and fiducial (dashed blue) pairs. The X-ray luminosity is retuned at each redshift, with $\log_{10}[L_{\rm X}/(\mathrm{erg\,s^{-1}}\,M_\odot^{-1}\,\mathrm{yr})] = 39.82$, $39.82$, $39.81$ and $39.76$ for the optimistic model and $38.85$, $38.83$, $38.80$ and $38.76$ for the fiducial one, leaving residuals below $1\,\mathrm{mK}$ in all cases. At $z = 16$, the IGM is barely heated in either channel and the ratio is dominated by the matching residual.}
    \label{fig:ps_matched}
\end{figure*}

In figure~\ref{fig:ps_matched} we show these matched comparisons at $z = 16$, $14$, $12$ and $10$. For the optimistic model, the median ratio over $0.05 < k < 0.8\,\mathrm{cMpc}^{-1}$ grows as heating proceeds, from $1.04$ at $z = 16$, through $1.21$ at $z = 14$ and $1.35$ at $z = 12$, to $1.55$ at $z = 10$: once the IGM is substantially heated, CR heating produces $\approx 55\%$ more 21-cm power than X-ray heating delivering the same mean signal. The fiducial model, which deposits an order of magnitude less energy, shows almost no excess, its median ratio remaining between $1.03$ and $1.06$ at all four redshifts. The power spectrum therefore separates the two channels only if CRs heat the IGM appreciably, i.e., for efficiencies above the Pop~II supernova budget (section~\ref{sec:escape}); at lower efficiencies, neither channel imprints appreciable temperature structure before reionization, and the two are degenerate in the power spectrum as well. We have no matched pair at the top of that budget, $3\times10^{48}\,\mathrm{erg}\,M_\odot^{-1}$, so the efficiency at which the signature appears is bracketed by the fiducial and optimistic pairs rather than measured. At $z = 16$, where $\langle x_{\rm HI} \rangle \approx 0.94$ and neither channel has appreciably heated the gas, the optimistic ratio scatters around unity, as the matching residual is comparable to the differences being measured. The matched comparison also clarifies the amplitude inversion discussed above: at fixed $\langle \delta T_b \rangle$, CR and X-ray heating differ in median power by at most $\approx 55\%$, so the inversion reflects the different mean contrast of the models rather than any property of CR transport.

The scale dependence of the ratio evolves together with its amplitude, and is the clearer of the two signatures. At $z = 14$, the optimistic ratio rises with wavenumber, from $0.87$ at $k = 0.05\,\mathrm{cMpc}^{-1}$ to $2.12$ at $k = 0.71\,\mathrm{cMpc}^{-1}$, a factor of $2.4$ across the resolved range, while the fiducial ratio in the same panel stays flat to within $10\%$. By $z = 12$, the rise is confined to the largest scales, with the ratio flat above $k \approx 0.1\,\mathrm{cMpc}^{-1}$, and by $z = 10$ the ratio is flat across the whole range, at $1.42$--$1.67$. The scale dependence is therefore a transient feature of the epoch in which heating is underway but incomplete, as expected from the different ranges of the two carriers. While the heated regions are still disjoint, the CR channel, whose contribution is cut off once the protons thermalize (section~\ref{sec:architecture}), adds power preferentially on the scales of the source distribution relative to the more penetrating X-rays; once heating has filled the volume, the two channels differ only in the overall contrast they produce, and the excess becomes scale independent. Because the transport approximation spreads each proton's heat over a larger region than it can physically reach, the scale dependence we find is likely a lower bound on the true signature. This scale dependence is qualitatively consistent with the high-wavenumber amplification found by GJ23.\footnote{A velocity-induced acoustic oscillation (VAO) feature~\citep{Munoz2019, Munoz2022}, sourced by the relative velocity between baryons and dark matter~\citep{TseliakhovichHirata2010}, is visible near $k \approx 0.16\,\mathrm{cMpc}^{-1}$ in all of our models. Its prominence relative to the local continuum agrees to within a few percent between the members of each matched pair, so the feature is inherited from the shared source field and does not discriminate between the two heating channels.}

The kinetic temperature offers a complementary view. Because the pairs are matched in $\langle \delta T_b \rangle$ rather than in $\langle T_{\rm K} \rangle$, the X-ray model requires a higher mean temperature to produce the same mean signal: by $5\%$ at $z = 16$, $15\%$ at $z = 14$, $27\%$ at $z = 12$ and $19\%$ at $z = 10$, where the optimistic pair has $\langle T_{\rm K} \rangle_{\rm HI} = 17.4$ and $20.7\,\mathrm{K}$, respectively. In other words, CR heating produces the same global signal as X-ray heating at a lower mean temperature, and more 21-cm power.

%For the pessimistic model no matched X-ray counterpart exists at $z = 10$: switching X-ray heating off entirely leaves $\langle \delta T_b \rangle = -156.38\,\mathrm{mK}$, below the $-151.54\,\mathrm{mK}$ of the CR model, so that the adiabatic floor is already colder than the CR model is able to reach. Compared against that floor the CR power is marginally lower, by $3$ to $7$\%, consistent with the small residual heating the CR channel supplies.

For efficient CR injection, the 21-cm power spectrum therefore retains differences that the global signal does not. In the global signal, the CR and X-ray troughs differ in depth and in position (by $\Delta z \approx 2$--$4$), differences that can be erased by retuning either efficiency. In the power spectrum, an excess in amplitude that grows to $\approx 55\%$ by $z = 10$ and, at $z \approx 14$, a rise of the ratio with wavenumber survive this retuning. The most discriminating epoch is therefore not the one in which the heating is strongest, but the one in which it is still underway, when the two channels differ in the shape of the power spectrum as well as in its amplitude.

\subsection{MCG-only injection and the flattened trough}\label{sec:direct_injection_results}

Thus far we have assumed that all star-forming halos inject CRs. We now turn to the MCG-only scenario of section~\ref{sec:direct_injection_physics}. GJ23 showed that, once the LW-elevated molecular-cooling threshold crosses $M_{\rm cr}^{\rm max}$, CR injection switches off, the IGM returns to adiabatic cooling, and $\langle T_{\rm K} \rangle(z)$ develops a local maximum followed by a secondary minimum, which for a limited range of $M_{\rm cr}^{\rm max}$ flattens the bottom of the absorption trough. Such a shape is reminiscent of the profile reported by EDGES, a flattened Gaussian centered at $78\,\mathrm{MHz}$ with an amplitude of $0.5\,\mathrm{K}$ \citep{Bowman2018}, although that detection has been questioned on modeling grounds \citep{Hills2018} and is in tension with the non-detection by SARAS~3 \citep{Singh2022}. We first scan $M_{\rm cr}^{\rm max}$ at fixed efficiency, and then ask whether a higher efficiency can recover the flattened trough; these runs include Ly$\alpha$ and CMB heating.

\begin{figure*}
    \centering
    \includegraphics[width=\textwidth]{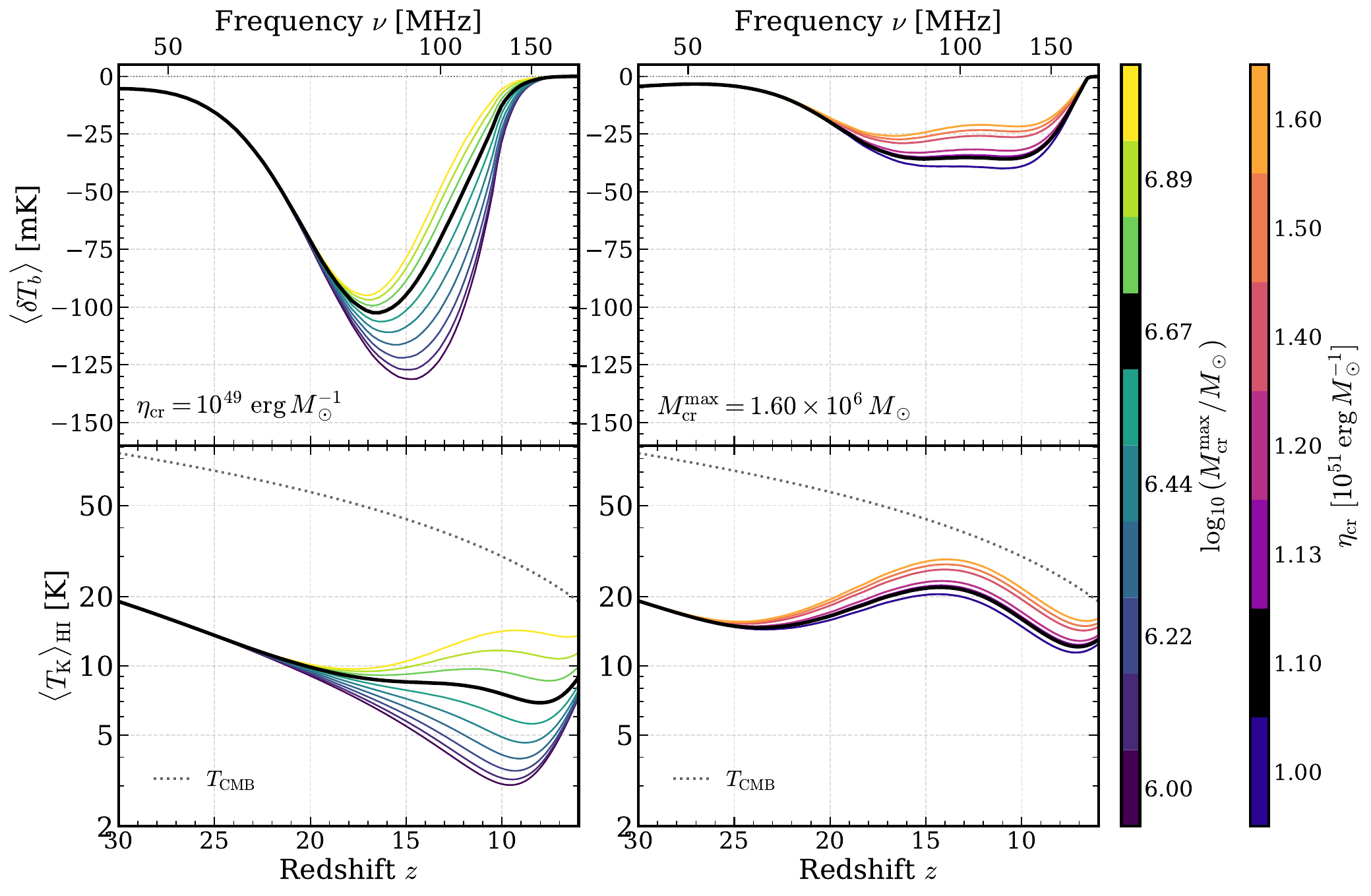}
    \caption{Global 21-cm signal (top row) and kinetic temperature of the neutral IGM (bottom row) for MCG-only CR heating, including Ly$\alpha$ and CMB heating. \textit{Left:} $\eta_{\rm cr} = 10^{49}\,\mathrm{erg}\,M_\odot^{-1}$, with $M_{\rm cr}^{\rm max}$ scanned logarithmically from $10^{6}$ to $10^{7}\,M_\odot$ (color bar, one band per run); the black curve marks the run closest to $M_{\rm cr}^{\rm max} = 4.57\times10^{6}\,M_\odot$, the value for which \citet{GesseyJones2023} reported a flat-bottomed trough, with Pop~III sources only, $\eta_{\rm cr} = 3\times10^{48}\,\mathrm{erg}\,M_\odot^{-1}$ and $\mathrm{d}N/\mathrm{d}E_{\rm K} \propto E_{\rm K}^{-2}$ between $10^{-3}$ and $10^{8}\,\mathrm{MeV}$ (their figure~10). \textit{Right:} $M_{\rm cr}^{\rm max} = 1.60\times10^{6}\,M_\odot$, with $\eta_{\rm cr} = (1.0$--$1.6)\times10^{51}\,\mathrm{erg}\,M_\odot^{-1}$ (color bar, one band per run); the black curve marks the flattest trough, at $\eta_{\rm cr} = 1.1\times10^{51}\,\mathrm{erg}\,M_\odot^{-1}$, flat to within $1\,\mathrm{mK}$. Dotted curves show $T_{\rm CMB}$.}
    \label{fig:minihalo}
\end{figure*}

We scan $M_{\rm cr}^{\rm max}$ logarithmically over $10^{6}$--$10^{7}\,M_\odot$, with $\eta_{\rm cr} = 10^{49}\,\mathrm{erg}\,M_\odot^{-1}$ and $\alpha = 4.2$ (left column of figure~\ref{fig:minihalo}). The kinetic temperature (bottom left) qualitatively reproduces the central thermal-history result of GJ23. For high $M_{\rm cr}^{\rm max}$, injection continues until late times, and $\langle T_{\rm K} \rangle$ develops a local maximum near $z \approx 9$--$11$, followed by a secondary minimum once injection stops. For low $M_{\rm cr}^{\rm max}$, injection shuts off before CR heating has raised $T_{\rm K}$ appreciably, and the curves approach the no-heating case. The shutoff mechanism thus operates for our injection spectrum as well. The global signal (top left), however, varies monotonically with $M_{\rm cr}^{\rm max}$, and no value in the grid flattens the trough, including the value $M_{\rm cr}^{\rm max} = 4.57\times10^{6}\,M_\odot$ highlighted by GJ23.

The reason is timing. Through the factor $(1 - T_{\rm CMB}/T_{\rm S})$ in eq.~\eqref{eq:dTb}, with $T_{\rm S} \simeq T_{\rm K}$, the 21-cm signal responds strongly to changes in $T_{\rm K}$ while the gas is much colder than the CMB, and progressively less as $T_{\rm K}$ approaches $T_{\rm CMB}$ and reionization reduces $x_{\rm HI}$. At $\eta_{\rm cr} = 10^{49}\,\mathrm{erg}\,M_\odot^{-1}$, the $T_{\rm K}$ feature forms only at $z \sim 10$, well after the trough minimum at $z \sim 15$, when the signal is already recovering: the feature is real, but it occurs too late to be seen in $\langle \delta T_b \rangle$.
%; it should still be visible in the cell-to-cell fluctuations of $T_K$, and hence in the power spectrum.

This suggests that the absence of flattening reflects the amplitude of CR heating rather than the shape of the SFRD evolution. A flattened trough requires two conditions: CR heating must raise $T_{\rm K}$ appreciably before injection stops, which is controlled by $\eta_{\rm cr}$, and the shutoff must occur within the trough, which is controlled by $M_{\rm cr}^{\rm max}$ through the LW history. As a larger $M_{\rm cr}^{\rm max}$ admits more sources, and thus requires a lower $\eta_{\rm cr}$ for the same heating, the two parameters are partly degenerate. We therefore run a grid in $(M_{\rm cr}^{\rm max}, \eta_{\rm cr})$ along this degeneracy, followed by successively finer grids around the flattest case.

A higher efficiency does recover the flattened trough. Flattening occurs along a narrow boundary in the $(M_{\rm cr}^{\rm max}, \eta_{\rm cr})$ plane, which separates two regimes. On one side, heating is too weak or stops too early, and the trough is deep, with a single late minimum. On the other, heating starts early enough to produce a shallow early minimum, after which the signal rises toward zero. Along the boundary, the two minima have nearly equal depths. The right column of figure~\ref{fig:minihalo} shows this transition at fixed $M_{\rm cr}^{\rm max}$: increasing $\eta_{\rm cr}$ moves the deeper minimum from the late to the early side of the trough, passing through a flat bottom in between. The flattest case, at $M_{\rm cr}^{\rm max} = 1.60\times10^{6}\,M_\odot$ and $\eta_{\rm cr} \simeq 1.1\times10^{51}\,\mathrm{erg}\,M_\odot^{-1}$, is flat to within $\sim 1\,\mathrm{mK}$ over $10 \lesssim z \lesssim 16$.

The kinetic temperature (bottom right of figure~\ref{fig:minihalo}) shows how the flat bottom forms: CR heating lifts $T_{\rm K}$ off the adiabatic track until injection switches off near $z \sim 14$, after which the gas cools again, though more slowly than adiabatic expansion alone would imply, remaining well below $T_{\rm CMB}$. After the shutoff, the cooling gas would deepen the trough again, but by then reionization has begun to reduce the neutral fraction, and the two effects nearly cancel until reionization drives the signal to zero. The redshift extent of the flat bottom therefore depends on the reionization history as well as on the CR parameters.

The required efficiency, however, lies beyond the supernova energy budget. As shown in section~\ref{sec:escape}, even if all Pop~III stellar mass formed in the most massive pair-instability progenitors, $\eta_{\rm cr} \lesssim 10^{50}\,\mathrm{erg}\,M_\odot^{-1}$, an order of magnitude below the value required here. Because MCG-only injection is sourced by the Pop~III SFRD, the quantity actually constrained is the product $\eta_{\rm cr} f_\star^{(\rm III)} \simeq 3.5\times10^{48}\,\mathrm{erg}\,M_\odot^{-1}$, for the EOS21 value $f_\star^{(\rm III)} = 10^{-2.5}$. For the higher Pop~III stellar fraction of GJ23, $f_\star^{(\rm III)} = 0.02$, the same product would correspond to $\eta_{\rm cr} \simeq 1.7\times10^{50}\,\mathrm{erg}\,M_\odot^{-1}$, still nearly a factor of two above the limit (this scaling is approximate, as the two galaxy models also differ in the shape of the SFRD). The flattened trough is therefore attainable only beyond the most extreme CR yield that Pop~III supernovae could supply. Even then, it does not explain the EDGES profile: the trough is $\sim 35\,\mathrm{mK}$ deep rather than $0.5\,\mathrm{K}$, and spans $10 \lesssim z \lesssim 16$ rather than being centered at $z \approx 17$, since CR heating, like any heating, can only make the trough shallower than in the adiabatic case. Within the energy budget, the LW-driven shutoff leaves a clear feature in $\langle T_{\rm K} \rangle$ but not in the global signal.

% ------------------------------------------------------------

% ============================================================
\section{Conclusions}\label{sec:conclusions}

Cosmic rays accelerated by the first supernovae are a plausible but poorly constrained source of heat for the early IGM, and their imprint on the 21-cm signal depends on how they are injected and how they deposit their energy. We have presented a new implementation of CR heating in \texttt{21cmFAST} that: (i) replaces the kinetic-energy power-law injection spectrum of GJ23 with the momentum power law predicted by diffusive shock acceleration, used previously only in calculations without spatially resolved 21-cm signals~\citep{Leite2017, JanaNath2018}; (ii) includes Coulomb losses off free electrons alongside the excitation, ionization and adiabatic losses; and (iii) sources the CRs from the data-informed populations of atomic-cooling and molecularly cooled galaxies of the EOS21 model.

We find that CR heating is about an order of magnitude weaker than previously estimated. Because 99\% of the heat comes from protons injected below $\approx 40$\,MeV, what matters is how much energy the spectrum places at these sub-relativistic energies. For our momentum power law with $\alpha = 4.2$, only 4.3\% of the injected energy lies below the $30$\,MeV boundary that GJ23 use, against 32\% for a kinetic-energy power law. Thus, at matched injection efficiency, the momentum power law alone deposits $\approx 12$ times less heat than the kinetic-energy power law at $z = 15$. Additionally, the more rapid SFRD evolution of the realistic galaxy model reduces the heating by a factor of $\approx 4$, and the two effects, which partly overlap, combine for a reduction of $\approx 28$.

At the efficiencies that supernovae can supply, CR heating is weaker than X-ray heating at the EOS21 fiducial luminosity. Even at $\eta_{\rm cr} = 10^{49}\,\mathrm{erg}\,M_\odot^{-1}$, above the Pop~II ceiling of $\approx 3\times10^{48}\,\mathrm{erg}\,M_\odot^{-1}$, CR-heated models remain cooler than the X-ray-heated one, and their absorption troughs lie later, by $\Delta z \approx 2$ at $10^{49}\mathrm{erg}\,M_\odot^{-1}$ and $ \Delta z \approx 4$ at injection efficiencies of $10^{48}\,\mathrm{erg}\,M_\odot^{-1}$. As this offset can be erased by retuning either efficiency, the global 21-cm signal alone cannot separate the two channels. Coulomb losses raise $\langle T_{\rm K} \rangle$ by $\lesssim 2\%$ and shift the global signal by $\lesssim 1.3$\,mK.

Distinguishing CR from X-ray heating with the 21-cm power spectrum requires CR efficiencies above those expected from local observations by a factor of a few, as bracketed by our runs at efficiencies of $10^{48}$ and $10^{49}\,\mathrm{erg}\,M_\odot^{-1}$, between which the top of the local range lies. Comparing the two channels at matched global signal, so that any difference reflects how the heat is distributed rather than the stage of heating the two channels have reached, we find that once heating is underway, CR heating produces more 21-cm power than X-ray heating. At $\eta_{\rm cr} = 10^{49}\,\mathrm{erg}\,M_\odot^{-1}$, about three times the top of the local range, the median excess grows from $21\%$ at $z = 14$ to $55\%$ at $z = 10$; at $10^{48}\,\mathrm{erg}\,M_\odot^{-1}$ it is only 3--6\%. At $z \sim 14$, while heating is underway but incomplete, the ratio also rises with wavenumber, by a factor of $2.4$ across the scales we resolve, a shape difference the global signal cannot access and one that disappears by $z = 10$, once heating has filled the volume. Because our light-travel transport of CRs smooths the heating field, this scale dependence is likely a lower bound on the true signature. Compared at their reference efficiencies, the two channels differ by far more, by up to two orders of magnitude, but most of that separation reflects the different stages of thermal evolution the two have reached rather than any property of the carriers themselves.

MCG-only injection, which previous work found could produce a flattened absorption trough reminiscent of the profile claimed by EDGES, also requires efficiencies well above the local range. The LW-driven shutoff of injection leaves a clear feature in the kinetic temperature but, at $\eta_{\rm cr} = 10^{49}\,\mathrm{erg}\,M_\odot^{-1}$, not in the global signal, because the feature forms only after the absorption minimum, once the signal is already recovering. A flattened absorption trough appears only for $\eta_{\rm cr} \approx 1.1\times10^{51}\,\mathrm{erg}\,M_\odot^{-1}$, over two orders of magnitude above the local range and beyond what even the most massive Pop~III supernovae could supply under generous assumptions about acceleration and escape. Even then, the trough is only $\sim 35$\,mK deep, far shallower than the EDGES profile.

The practical consequence is that upper limits on the 21-cm power spectrum, such as those from HERA~\citep{HERA2023, HERA2025}, should constrain the CR channel mainly through the resulting thermal history. At fiducial values, the canonical heating channel by X-rays from HMXBs is significantly more efficient compared to CR heating; as a result, current interpretation of 21cm upper limits is unlikely to be impacted by ignoring CR heating. At a fixed thermal history separating the two channels would require measurements at several wavenumbers during the heating transition, or the combination of the power spectrum with the global signal. Even so, unless early supernovae were several times more efficient CR sources than local ones, 21-cm power spectra measurements can hardly tell CR heating from X-ray heating. We defer a joint inference of CR and X-ray heating from current HERA upper limits to future work.

% ============================================================
\acknowledgments

We thank Thomas Gessey-Jones for helpful correspondence regarding the parameters used in \citet{GesseyJones2023} and for providing supplementary data. 
Claude Opus 5.5 was used to assist with language revision and verification of the analysis workflow. All scientific decisions, validation of the results, and responsibility for the final manuscript remain with the authors.

% ============================================================
\section*{Data availability}

Our code is publicly available at \url{https://github.com/michaelpaganocosmo/21cmFAST-CosmicRays}.

% ============================================================

% ============================================================

% ============================================================
% References: inline bibliography, ordered by first citation (JCAP numeric style).
% SpitzerScott1969 was added in this format; verify its title on ADS.

\label{lastpage}
\end{document}